\documentclass[fleqn,usenatbib]{mnras}

\usepackage[T1]{fontenc}
\usepackage{ae,aecompl}

\usepackage{graphicx}	
\usepackage{adjustbox}  
\usepackage{amsmath}	
\usepackage{amssymb}	
\usepackage{bm}         
\usepackage{txfonts}
\usepackage{etoolbox}
\usepackage{float}
\usepackage{enumerate}
\makeatletter 
  \patchcmd{\NAT@citex}
    {\@citea\NAT@hyper@{%
      \NAT@nmfmt{\NAT@nm}%
      \hyper@natlinkbreak{\NAT@aysep\NAT@spacechar}{\@citeb\@extra@b@citeb}%
      \NAT@date}}
    {\@citea\NAT@nmfmt{\NAT@nm}%
    \NAT@aysep\NAT@spacechar\NAT@hyper@{\NAT@date}}{}{}

  \patchcmd{\NAT@citex}
    {\@citea\NAT@hyper@{%
      \NAT@nmfmt{\NAT@nm}%
      \hyper@natlinkbreak{\NAT@spacechar\NAT@@open\if*#1*\else#1\NAT@spacechar\fi}%
        {\@citeb\@extra@b@citeb}%
      \NAT@date}}
    {\@citea\NAT@nmfmt{\NAT@nm}%
    \NAT@spacechar\NAT@@open\if*#1*\else#1\NAT@spacechar\fi\NAT@hyper@{\NAT@date}}
    {}{}
\makeatother

\usepackage{bm}
\usepackage[utf8]{inputenc}
\usepackage{ulem}

\newcommand\orcid[1]{\href{http://orcid.org/#1}{\adjustbox{trim={-.15\width} {0\height} {-.15\width} {0\height},clip}{\includegraphics[height=12pt]{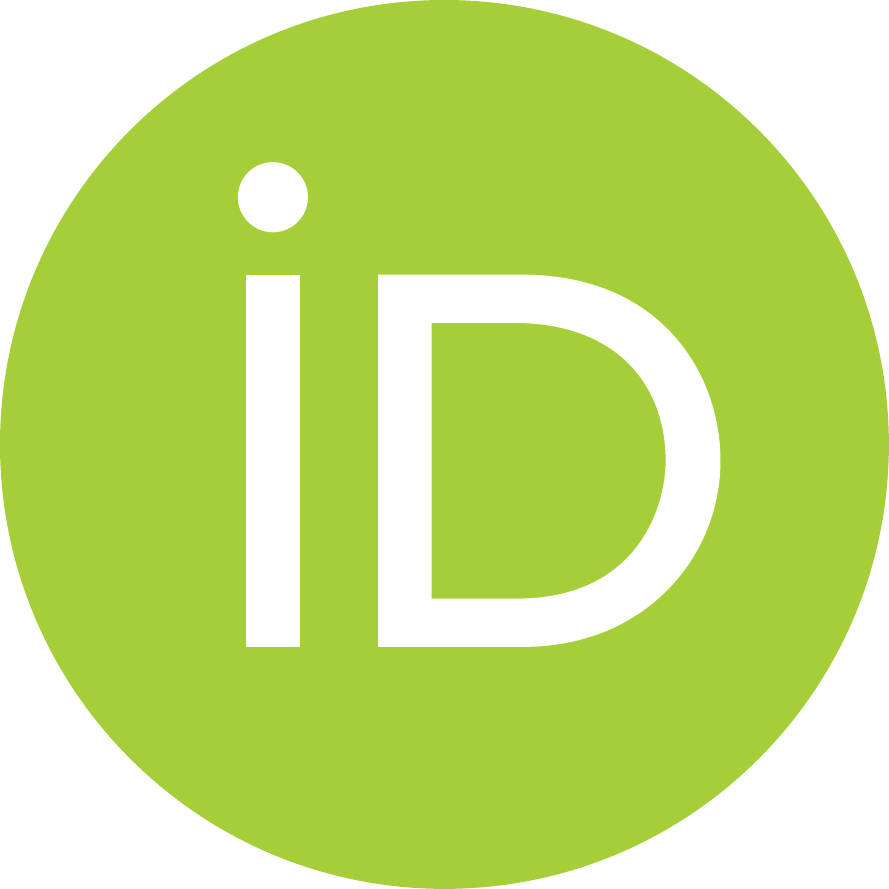}}}}

\title[GMCs with different density profiles]{Effects of initial density profiles on massive star cluster formation in giant molecular clouds}

\author[Y. Chen, H. Li, and M. Vogelsberger]{Yingtian Chen$^{1,2}$\orcid{0000-0002-5970-2563}, 
  Hui Li$^{1}$\orcid{0000-0002-1253-2763}\thanks{E-mail: \href{mailto:hliastro@mit.edu}{hliastro@mit.edu}}\thanks{NHFP Hubble Fellow}
  and Mark Vogelsberger$^{1}$\orcid{0000-0001-8593-7692}
  \\
$^{1}$Department of Physics, Kavli Institute for Astrophysics and Space Research, Massachusetts Institute of Technology, Cambridge, MA 02139, USA\\
$^{2}$School of Physics, Peking University, Beijing 100871, China
}

\date{Accepted XXX. Received YYY; in original form ZZZ}

\pubyear{2020}

\begin{document}
\label{firstpage}
\pagerange{\pageref{firstpage}--\pageref{lastpage}}
\maketitle

\begin{abstract}
We perform a suite of hydrodynamic simulations to investigate how initial density profiles of giant molecular clouds (GMCs) affect their subsequent evolution. We find that the star formation duration and integrated star formation efficiency (SFE) of the whole clouds are not sensitive to the choice of different profiles but are mainly controlled by the interplay between gravitational collapse and stellar feedback. Despite this similarity, GMCs with different profiles show dramatically different modes of star formation. For shallower profiles, GMCs first fragment into many self-gravitation cores and form sub-clusters that distributed throughout the entire clouds. These sub-clusters are later assembled ``hierarchically'' to central clusters.  In contrast, for steeper profiles, a massive cluster is quickly formed at the center of the cloud and then gradually grows its mass via gas accretion. Consequently, central clusters emerged from clouds with shallower profiles are less massive and show less rotation than those with the steeper profiles. This is because 1) a significant fraction of mass and angular momentum in shallower profiles is stored in the orbital motion of the sub-clusters that are not able to merge into the central clusters 2) frequent hierarchical mergers in the shallower profiles lead to further losses of mass and angular momentum via violent relaxation and tidal disruption. Encouragingly, the degree of cluster rotations in steeper profiles is consistent with recent observations of young and intermediate-age clusters. We speculate that rotating globular clusters are likely formed via an ``accretion'' mode from centrally-concentrated clouds in the early Universe.
\end{abstract}




\section{Introduction}
\label{sec:introduction}

Giant molecular clouds (GMCs) are the cradles of star clusters \citep{shu_star_1987, scoville_far-infrared_1989, mckee_theory_2007, krumholz_star_2019}, in which most stars form \citep[see][]{lada_embedded_2003}.
Understanding the modes and the efficiency with which stars and clusters form is one of the key problems in modern astrophysics.
Although numerous observational and theoretical efforts have been made to investigate GMC evolution \citep[see a recent review,][]{girichidis_physical_2020}, the complexity of the physical processes involved and the highly uncertain initial conditions imply that many aspects of the problem are still active areas of research.

One seemingly simple question is how efficiently GMCs form stars.
Previous work has investigated the star formation efficiency (SFE) of GMCs with different turbulence properties \citep[e.g.][]{krumholz_general_2005, padoan_simple_2012} and channels of stellar feedback \citep[e.g.][]{dale_ionizing_2012, walch_dispersal_2012, rogers_feedback_2013, dale_before_2014, myers_star_2014, geen_photoionization_2015, kim_modeling_2017, lucas_supernova_2020}.
Encouragingly, recent analytical and numerical studies have reached a general consensus that the integrated SFE of an individual cloud depends on the interplay between gravitational collapse and the strength of stellar feedback from massive stars \citep[e.g.][]{fall_stellar_2010, murray_disruption_2010, burkert_dependence_2013, raskutti_numerical_2016, grudic_when_2018}.
In our recent work \citep[][hereafter L19]{li_disruption_2019}, we surveyed a large range of GMC properties and stellar feedback strengths, and concluded that SFE depends strongly on the gas surface density of the clouds and the feedback momentum.
Similar models have been used to reproduce the large scatter in observed star formation efficiencies on cloud scales \citep[e.g.][]{grudic_nature_2019}.

However, most of the simulations mentioned above start from isolated turbulent clouds with a specific initial gas density distribution.
Therefore, whether the conclusions drawn from these simulations are sensitive to the choice of initial gas distribution or not is largely unknown.
Moreover, different simulations adopt dramatically different star formation and stellar feedback prescriptions, making it impractical to conduct useful comparisons on the effects of initial conditions on star formation.
Despite the over-simplified setup of the initial conditions in simulations, GMCs and molecular clouds in observations are indisputably complex. GMCs in the nearby universe show diverse gas profiles and probability distributions \citep{lombardi_2mass_2008, pirogov_density_2009, kainulainen_probing_2009, schneider_what_2013, schneider_understanding_2015, naranjo-romero_hierarchical_2015}.
Therefore, it is crucial to design control experiments that systematically investigate how the initial gas density distribution affects the subsequent gas evolution, star formation activities, and, eventually, the properties of star clusters.

A few studies have tried to explore the problem from this perspective.
For example, in a series of papers, \citet{girichidis_importance_2011, girichidis_importance_2012, girichidis_importance_2012-1} systematically investigated cloud evolution and star formation for different initial density profiles.
They found that different density profiles produced dramatically different initial mass functions: clouds with flat cores produce many low-mass stars, while concentrated density profiles tend to form one massive star particle.
However, their studies focused on $\sim0.1\mathrm{pc}$ molecular cores, rather than the GMCs where massive clusters emerge from.
Moreover, they did not consider the feedback from massive stars, which is crucial for understanding the efficiency of star formation.

In this work, we perform a suite of hydrodynamic simulations to systematically evaluate the impact of initial density profiles of GMCs on the formation and evolution of GMCs and massive star clusters. Following the numerical setup of L19, we design three initial conditions with different power-laws density profiles: $\rho(r)\propto r^0$, $\rho(r)\propto r^{-1.5}$, and $\rho(r)\propto r^{-2}$.
Our goal is to answer the following questions: 1) how does the initial density profile affect the timescale and efficiency of star formation in GMCs? 2) how does it control the nature of gas fragmentation and the assembly history of massive star clusters? 3) are there any systematic differences in the properties of massive star clusters that emerge from GMCs with different density profiles?

The rest of this paper is organized as follows.
Sec.~\ref{sec:methods} describes the numerical implementations in the moving-mesh code \textsc{arepo} and the setup of the initial conditions of GMCs.
In Sec.~\ref{sec:results}, we show the similarities and differences of the star formation history, gas accretion profiles, and dynamics of star clusters for three different density profiles. We find that the star formation duration and integrated SFE of the whole cloud depend weakly on initial profiles. In contrast, the fragmentation of the clouds and assembly history of star clusters are sensitive to the choice of profiles: GMCs with steeper initial density profiles show more centrally-concentrated star formation activities, while clouds with shallower profiles form star clusters more hierarchically with frequent mergers among sub-clusters. Central clusters emerged from GMCs with steeper profiles are more massive and have higher specific angular momentum than those formed in clouds with shallower profiles.
In Sec.~\ref{sec:discussion}, we discuss the physical origins and implications of our results. We summarize our conclusions in Sec.~\ref{sec:summary}.

\section{Methods}
\label{sec:methods}
The numerical setup of this work is similar to L19, except for the initial conditions.
Here, we briefly recap the physical models and key parameters of the simulations, and highlight the setup of the initial conditions of the GMCs with different density profiles.

\subsection{Numerical setup}
\label{sec:numericalSetup}

All simulations in this work are performed with \textsc{Arepo} \citep{springel_e_2010}, a moving-mesh, finite-volume hydrodynamic code employing a second-order unsplit Godunov scheme.
The control volumes are discretized by a Voronoi tessellation, which is generated from its dual Delaunay tessellation determined by a set of mesh-generating points.
Following L19, we include self-gravity, radiative cooling/heating, star formation, and momentum stellar feedback sub-grid models.
We employ an adaptive softening scheme for gas cells so that gravitational interactions are resolved down to the size of individual cells.
We allow gas cells to refine and de-refine in a quasi-Lagrangian fashion: the mass of all gas cells is around the target mass of $0.38\ M_\odot$ determined by the initial conditions, see Sec.~\ref{sec:initialConditions}.

We model the star formation process in a stochastic manner, where stellar particles are converted probabilistically from eligible cells, which are defined as cold, contracting, self-gravitating, and sufficiently dense cells.
The star formation density threshold is set to be $n_{\mathrm{cell}}=10^6\ \mathrm{cm^{-3}}$.
We vary this value from $10^4$ to $10^8\ \mathrm{cm^{-3}}$ and find that the GMC evolution and star formation activities do not depend on the choice of the threshold, see Appendix~\ref{append:threshold}.
The star particles are modelled as collisionless particles, whose Plummer-equivalent softening length is fixed to $4\times 10^{-3}\ \mathrm{pc}$. 
We employ a simple stellar feedback prescription by depositing mass and momentum from each stellar particle to their 32 nearest neighboring gas cells in a solid angle-weighted fashion.
The mass and momentum deposition rates are calculated based on a Kroupa initial mass function (IMF)-averaged values \citep{kroupa_variation_2001} for stellar winds, and the momentum feedback intensity is controlled by a normalization factor $f_{\mathrm{boost}}$, which is set to be $f_{\mathrm{boost}}=2$.

\begin{figure}
	\includegraphics[width=\columnwidth]{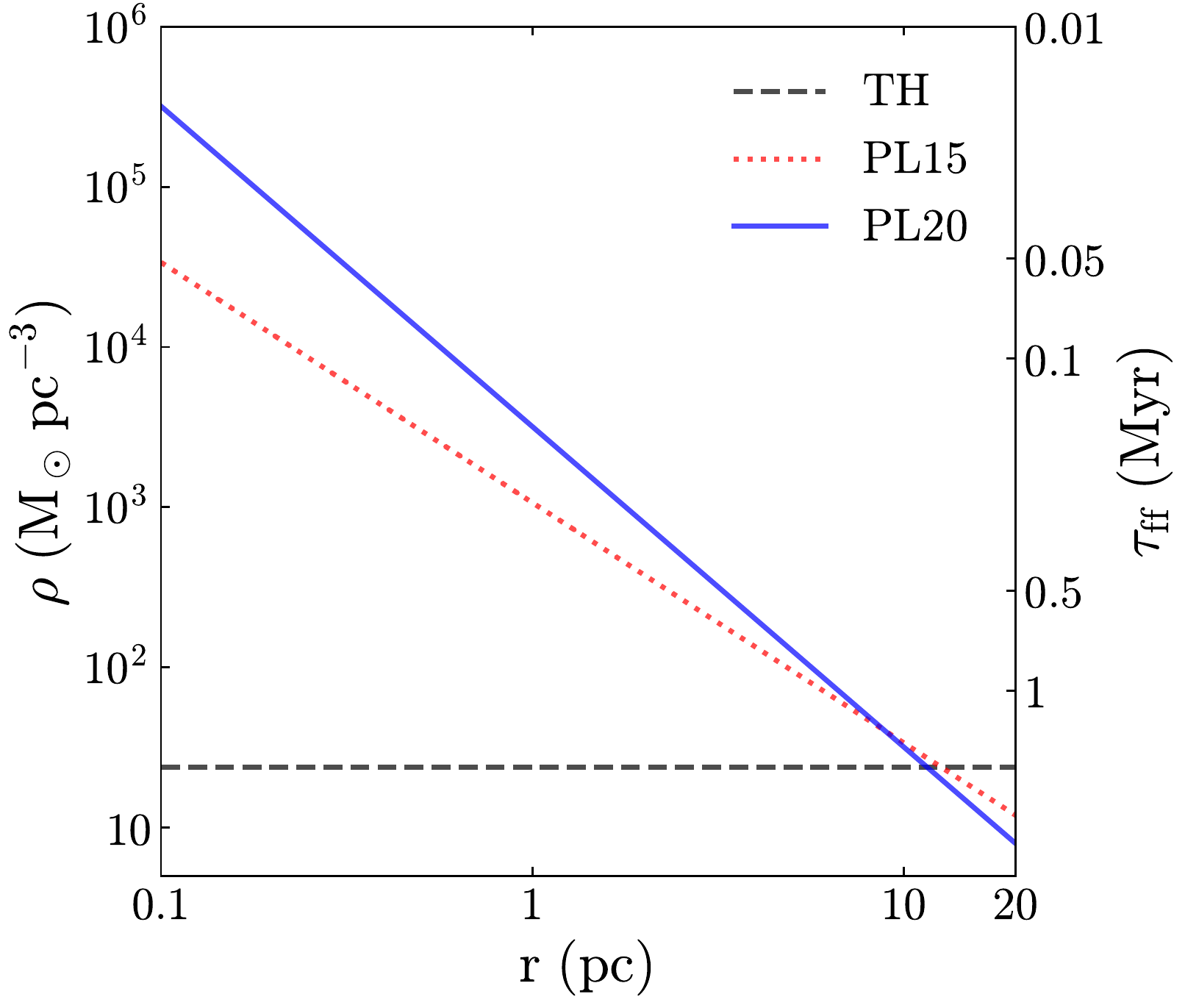}
    \caption{Initial density (left y-axis) and free-fall time (right y-axis) profiles of the TH (black dashed), PL15 (red dotted), and PL20 (blue solid) cases within the radius of $0.1-20\ \mathrm{pc}$.}
    \label{fig:iniPro}
\end{figure}

\begin{table*}
	\centering
	\caption{Summary of simulation results.}
	\label{tab:sum}
	\begin{tabular}{lccccccccc}
		\hline
		& $R_\mathrm{half}/R_\mathrm{GMC}$ $^a$ & $\tau_\mathrm{half}$ $^b$ & $\epsilon_\mathrm{int}$ $^c$ & $\tau_\mathrm{dur}$ $^d$ & $M_\mathrm{central}/M_\mathrm{GMC}$ $^e$ & $j_\mathrm{central}$ $^f$ & $E^\mathrm{rot}_\mathrm{central}/E^\mathrm{kin}_\mathrm{central}$ $^g$ & $f(\epsilon>0.5)$ $^h$ & $\alpha_\mathrm{central}$ $^i$ \\
		&  & $\rm{(Myr)}$ &  & $\rm{(Myr)}$ &  & $\rm{(pc\,km\,s^{-1})}$ &  &  & \\
		\hline
		TH & $0.79$ & $1.66$ & $0.542\pm 0.040$ & $1.07\pm 0.12$ & $0.078\pm 0.040$ & $1.24\pm 0.58$ & $0.181 \pm 0.117$ & $0.379\pm 0.099$ & $0.705 \pm 0.060$ \\
		PL15 & $0.63$ & $1.66$ & $0.501\pm 0.050$ & $1.17\pm 0.14$ & $0.180\pm 0.057$ & $1.89\pm 0.31$ & $0.209 \pm 0.062$ & $0.401\pm 0.053$ & $0.768 \pm 0.033$ \\
		PL20 & $0.5$ & $1.43$ & $0.509\pm 0.043$ & $1.04\pm 0.14$ & $0.303\pm 0.046$ & $2.82\pm 0.83$ & $ 0.296 \pm 0.089$ & $0.461\pm 0.067$ & $0.786 \pm 0.037$ \\
		\hline
		\multicolumn{10}{l}{\textbf{Note.} Mean values and standard deviations are calculated from ten random seeds.}\\
		\multicolumn{10}{l}{$^a$ Fractions of half-mass radii to the entire radii of GMCs.}\\
		\multicolumn{10}{l}{$^b$ Half-mass time scales of GMCs, i.e., the free-fall times at half-mass radii.}\\
		\multicolumn{10}{l}{$^c$ Integrated star formation efficiencies of GMCs.}\\
		\multicolumn{10}{l}{$^d$ Star formation duration of GMCs, defined as the timescale during which the clouds form the central $80\%$ of their stars: $\tau_\mathrm{dur}=t_\mathrm{90}-t_\mathrm{10}$.}\\
		\multicolumn{10}{l}{$^e$ Mass fractions of final central clusters to the original clouds.}\\
		\multicolumn{10}{l}{$^f$ Specific angular momentum of central clusters.}\\
		\multicolumn{10}{l}{$^g$ Ratios of central clusters' rotational energies to their total kinetic energies.}\\
		\multicolumn{10}{l}{$^h$ Proportions of $\epsilon>0.5$ stellar particles in central clusters, where $\epsilon$ is the circularity described in Sec.~\ref{sec:dynamics&Substructure}.}\\
		\multicolumn{10}{l}{$^i$ Virial parameters of central clusters.}\\
	\end{tabular}
\end{table*}

\subsection{Initial conditions}
\label{sec:initialConditions}

To systematically study the impact of the initial density profiles of GMCs on massive star cluster formation, we construct three sets of initial conditions with different power-law density profiles and velocity fields that are mixtures of supersonic turbulence and rigid rotation.
The GMCs have an initially spherical shape with a radius of $R_{\mathrm{GMC}}=20\ \mathrm{pc}$ and a total mass of $M_{\mathrm{GMC}}=8\times10^5\ \mathrm{M_\odot}$.
We initialize each GMC with $N_{\mathrm{cell}}=128^3$ Voronoi cells of identical mass, which leads to a mass resolution of $m_\mathrm{res}=M_{\mathrm{GMC}}/N_{\mathrm{cell}}=0.38\ M_\odot$.

\subsubsection{Initial density profiles}
\label{sec:initialDensityProfile}

Motivated by the large range of density distribution of GMCs from observations and theoretical expectations, we study three different power-law density profiles.
First, we employ a uniform-density profile, $\rho(r)\propto r^0$ (top-hat, or TH), which is the simplest and most commonly used profiles in previous simulations \citep[e.g.][L19]{ostriker_density_2001, bonnell_hierarchical_2003, padoan_simple_2012, raskutti_numerical_2016, skinner_numerical_2015, mapelli_rotation_2017, grudic_nature_2019, kim_modeling_2019}.
Then we employ two centrally-concentrated gas density profiles $\rho(r)\propto r^{-1.5}$ (PL15) and $\rho(r)\propto r^{-2}$ (PL20), motivated by recent observations \citep[e.g.][]{mueller_physical_2002, pirogov_density_2009, palau_fragmentation_2014, schneider_understanding_2015, wyrowski_infall_2016, csengeri_alma_2017}. Theoretically, PL15 is suggested to be a natural consequence of self-gravitating turbulent flows \citep[e.g.][]{murray_star_2015}, while PL20 is consistent with the self-similar solution of scale-free gravitational collapse \citep[e.g.][]{larson_numerical_1969, penston_dynamics_1969, naranjo-romero_hierarchical_2015, donkov_density_2018, li_scale-free_2018}.

Fig.~\ref{fig:iniPro} shows the analytical shape of the three profiles as well as the corresponding free-fall timescales, $\tau_{\mathrm{ff}}(r)=[3\pi/32G\rho(r)]^{1/2}$, at different radii.
The TH case has a uniform density of $\sim24\ \mathrm{M_\odot\,pc^{-3}}$, while the density ranges for the PL15 and PL20 cases are $1.2\times10^1-3.4\times10^4\ \mathrm{M_\odot\,pc^{-3}}$ and $8\times10^0-3.2\times10^5\ \mathrm{M_\odot\,pc^{-3}}$, respectively, within $0.1-20\ \mathrm{pc}$.
As expected, the free-fall timescales of the PL15 and PL20 cases are considerably shorter near the center of the clouds compared to that of the TH case.
To characterize the overall GMC evolution, we introduce the half-mass timescale $\tau_{\rm half}$, defined as the free-fall timescale at the half-mass radius $R_{\rm half}$.
The values of both $R_{\rm half}$ and $\tau_{\rm half}$ for the three profiles are listed in Tab.~\ref{tab:sum}.
We emphasize that, although the density distribution is dramatically different, both $R_{\rm half}$ and $\tau_{\rm half}$ have similar values for all three profiles: $R_{\rm half}\sim0.5-0.79\ \mathrm{pc}$ and $\tau_{\rm half}\sim1.43-1.66\ \mathrm{Myr}$, respectively.

\subsubsection{Initial velocity}
\label{sec:initialVelocity}

We follow the ``T'' runs in L19 to set the initial velocity field as a mixture of turbulence and rigid rotation.
We initialize the turbulent field as a Gaussian random field in wave-number space with a power spectrum of $P(k)\propto k^{-4}$, which resembles the turbulence properties of GMCs \citep{dobbs_formation_2014}.
Then, we perform a Fourier transform to reconstruct the turbulent field in real space.
Performing a Fourier transfer on the TH case is computationally cheap, since the cell sizes are similar across the whole cloud and a small dynamical range of wavelength is required.
For the PL15 and PL20 cases, it is much more computationally expensive, since the gas cells near the cloud center are much smaller than those at large radii.
To alleviate this problem, we employ a multi-threading non-uniform FFT code \textsc{FINUFFT} \citep{barnett_parallel_2018}, which applies a novel ``exponential of semicircle'' kernel, as a substitute of the traditional FFT. To make sure that our results do not depend on the specific choice of the random seed, we generate ten different turbulent velocity fields by using different random seeds for each initial density profile.

We treat the overall velocity field as the linear combination of above turbulence and a rigid rotation around the $z$-axis, and this linear combination is normalized so that the clouds are initially in virial equilibrium, i.e.,
\begin{equation}
    \alpha=\frac{2E^{\mathrm{kin}}}{|E^{\mathrm{grav}}|}=\frac{2E_\mathrm{T}^{\mathrm{kin}}}{|E^{\mathrm{grav}}|}+\frac{2E_\mathrm{R}^{\mathrm{kin}}}{|E^{\mathrm{grav}}|}=\alpha_{\mathrm{T}}+\alpha_{\mathrm{R}}=1\:.
    \label{eq:virialEquilibrium}
\end{equation}
We follow L19 and set the virial parameters of turbulence and rotation as $\alpha_{\mathrm{T}}=0.9$ and $\alpha_{\mathrm{R}}=0.1$. 
As described in Sec.~\ref{sec:physicalOriginAM}, we will explore the physical origin of star cluster rotation.
To investigate whether the star cluster rotation is directly linked to the rotation of the initial conditions, we generate an additional suite of initial conditions that only contains a turbulence velocity field with zero initial angular momentum.

\begin{figure*}
	\includegraphics[width=0.66\columnwidth]{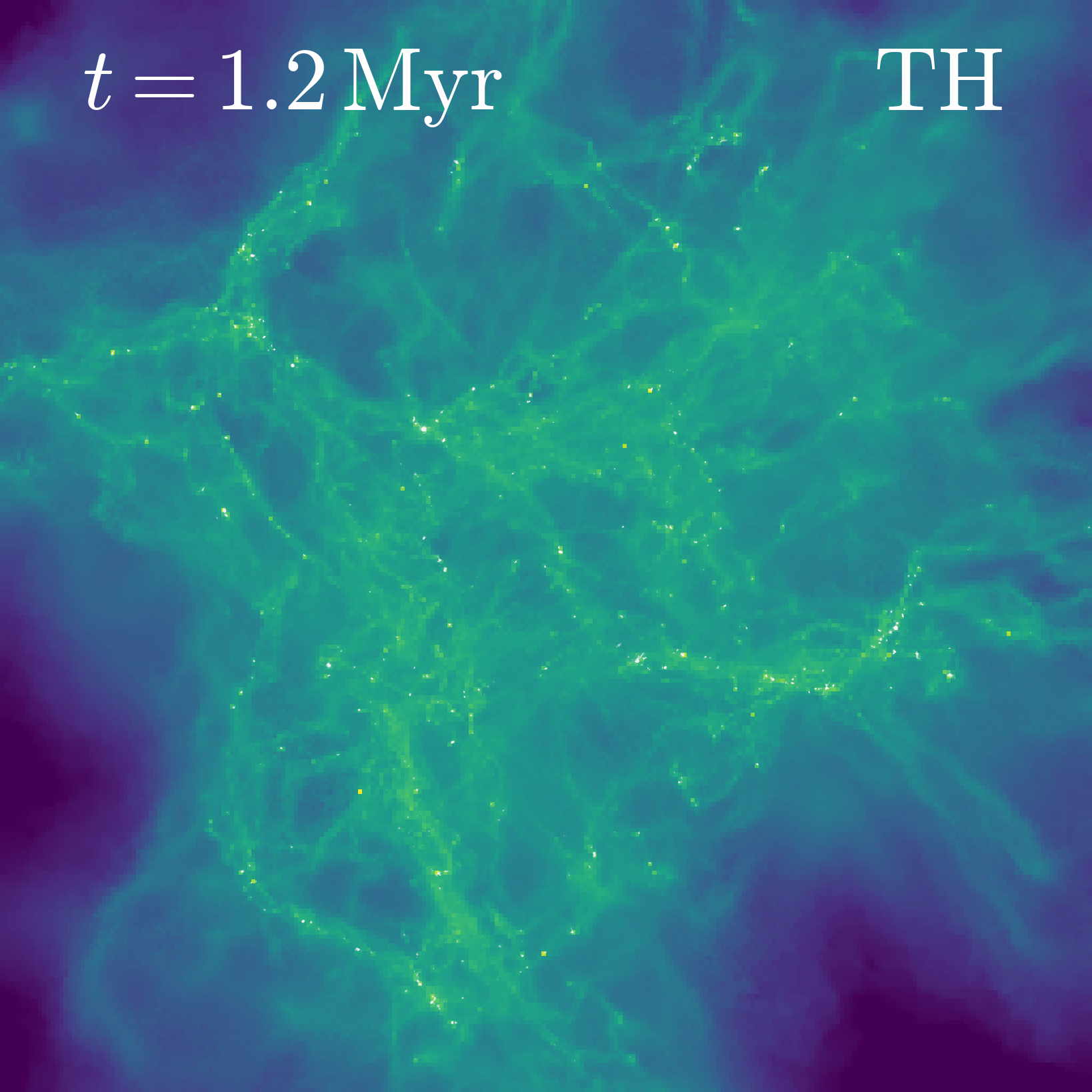}
	\hspace{0.001\columnwidth}
	\includegraphics[width=0.66\columnwidth]{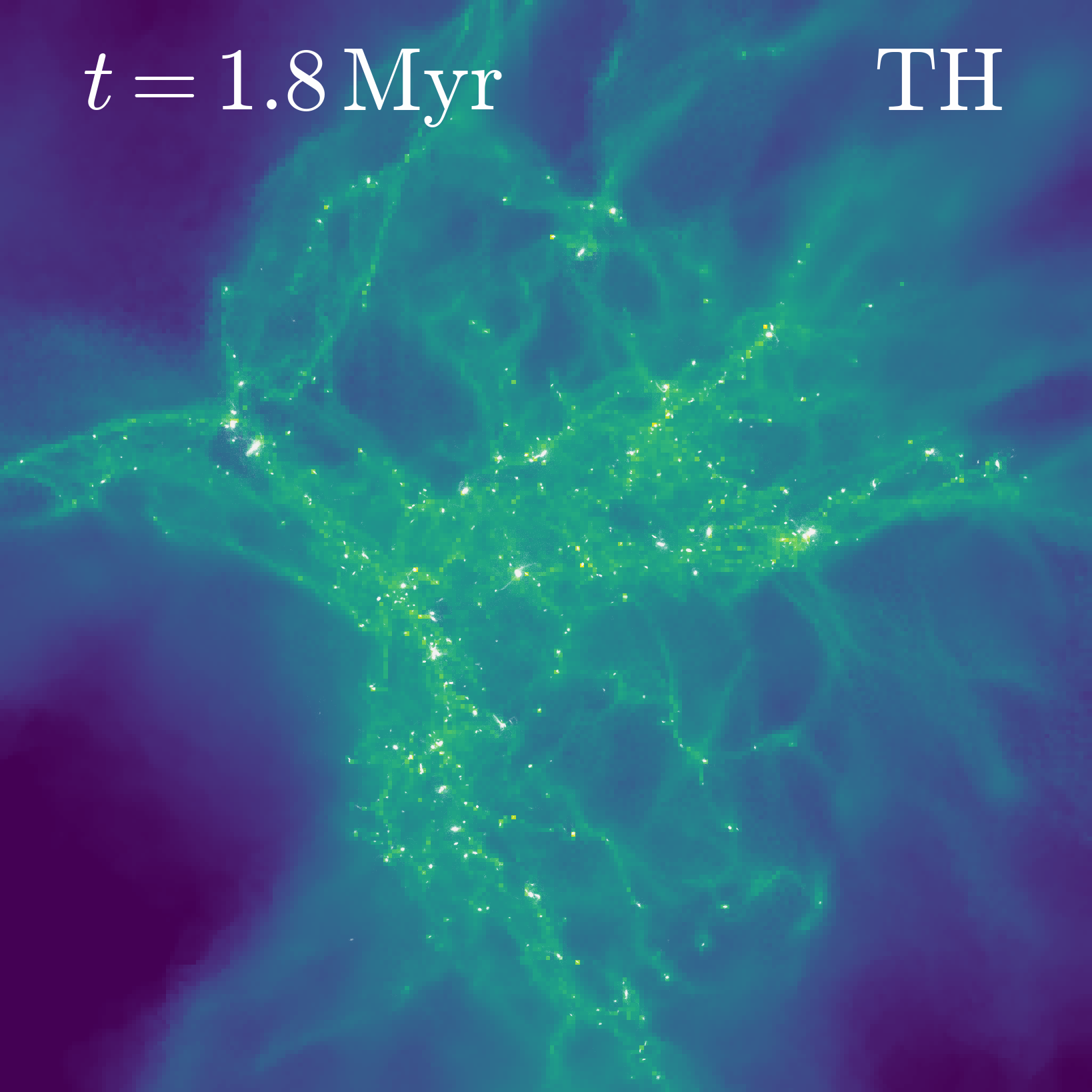}
	\hspace{0.001\columnwidth}
	\includegraphics[width=0.66\columnwidth]{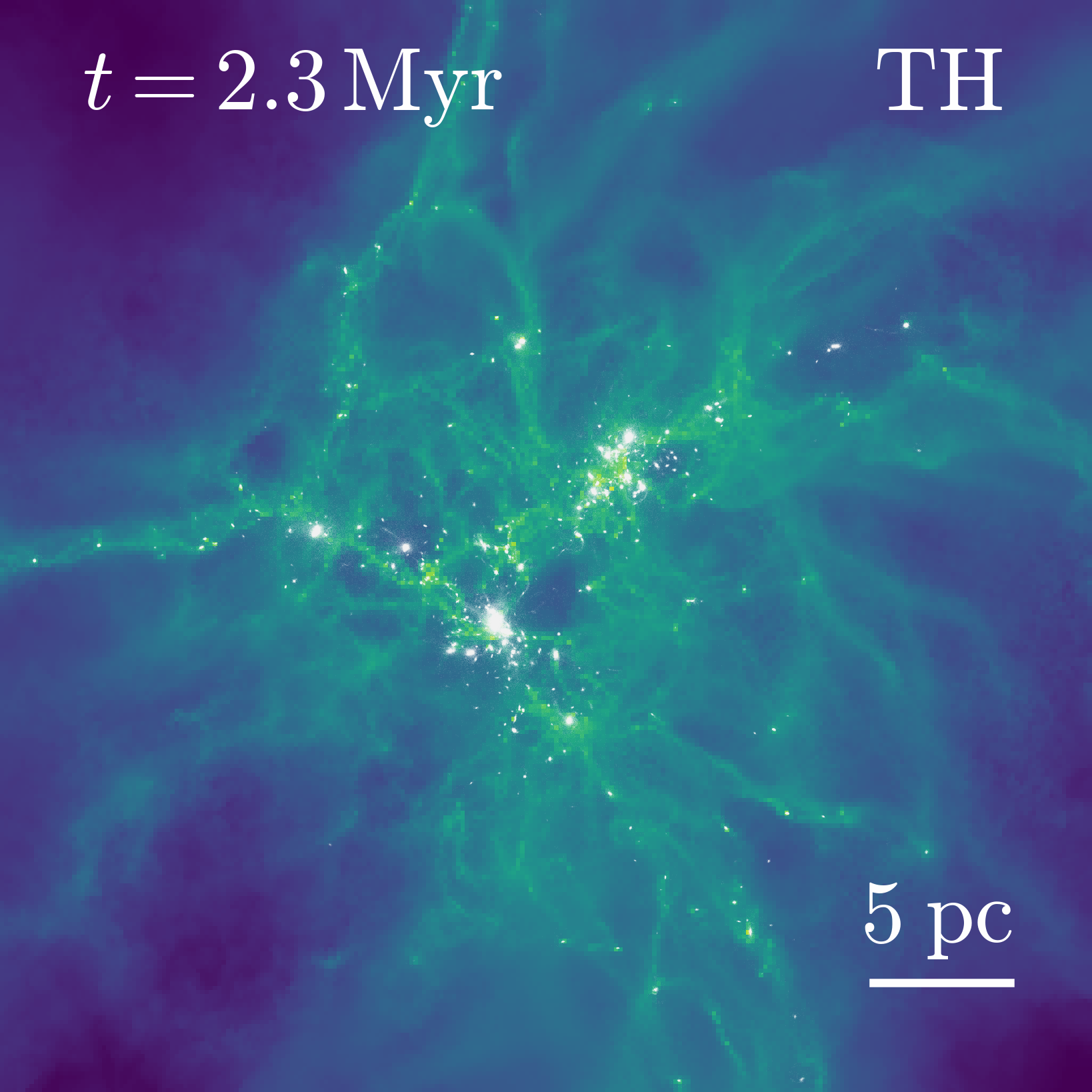}
	\vspace{0.01\columnwidth} \\
	\includegraphics[width=0.66\columnwidth]{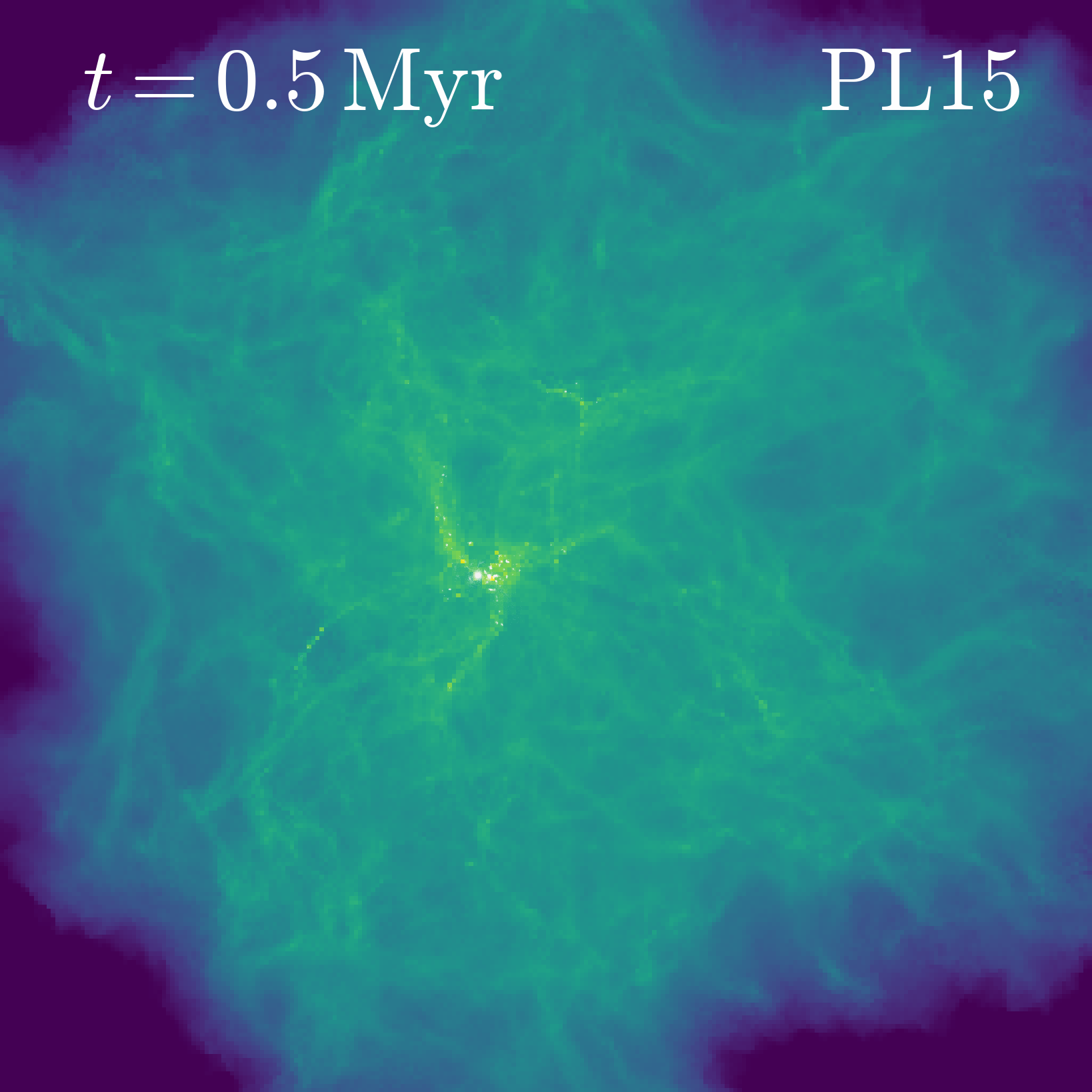}
	\hspace{0.001\columnwidth}
	\includegraphics[width=0.66\columnwidth]{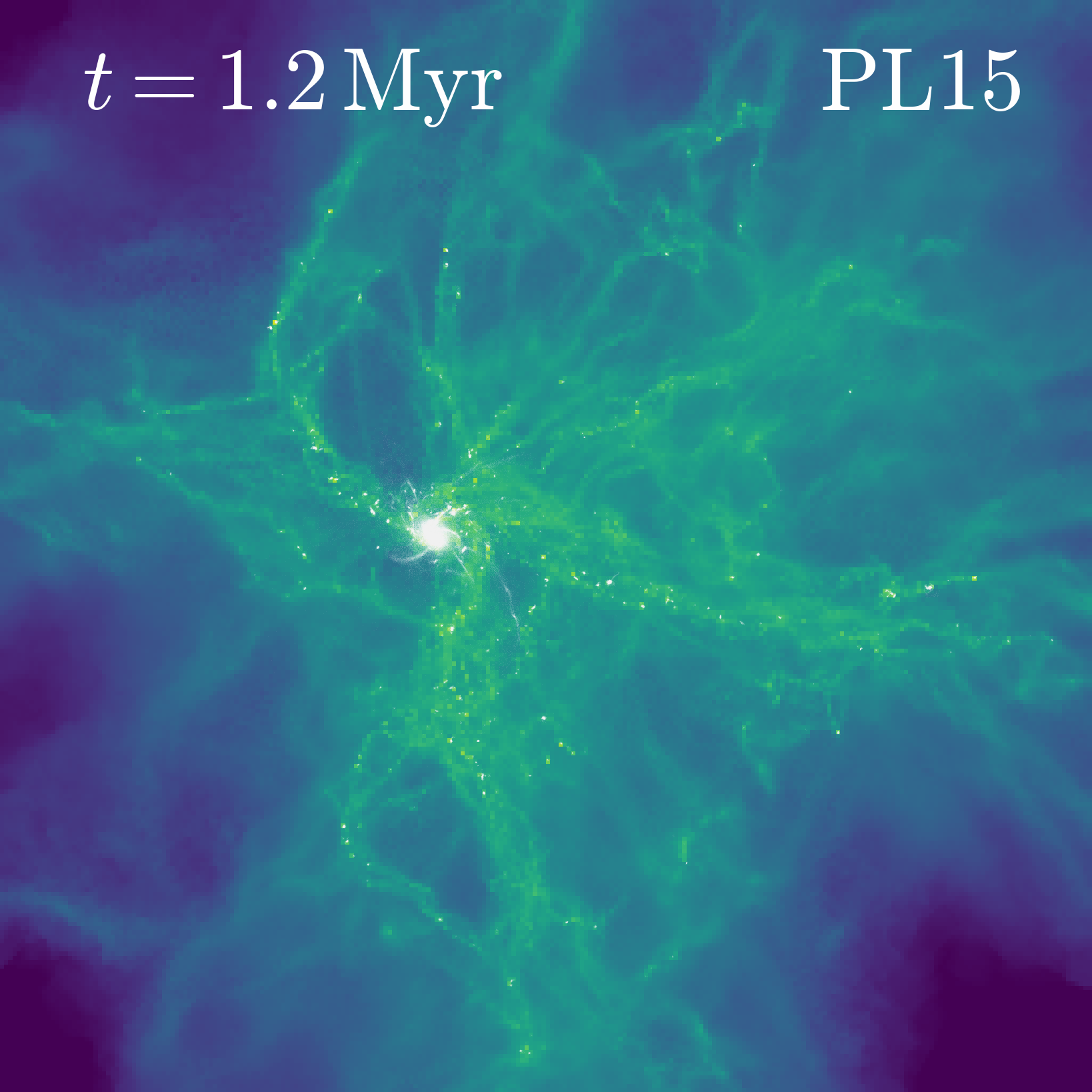}
	\hspace{0.001\columnwidth}
	\includegraphics[width=0.66\columnwidth]{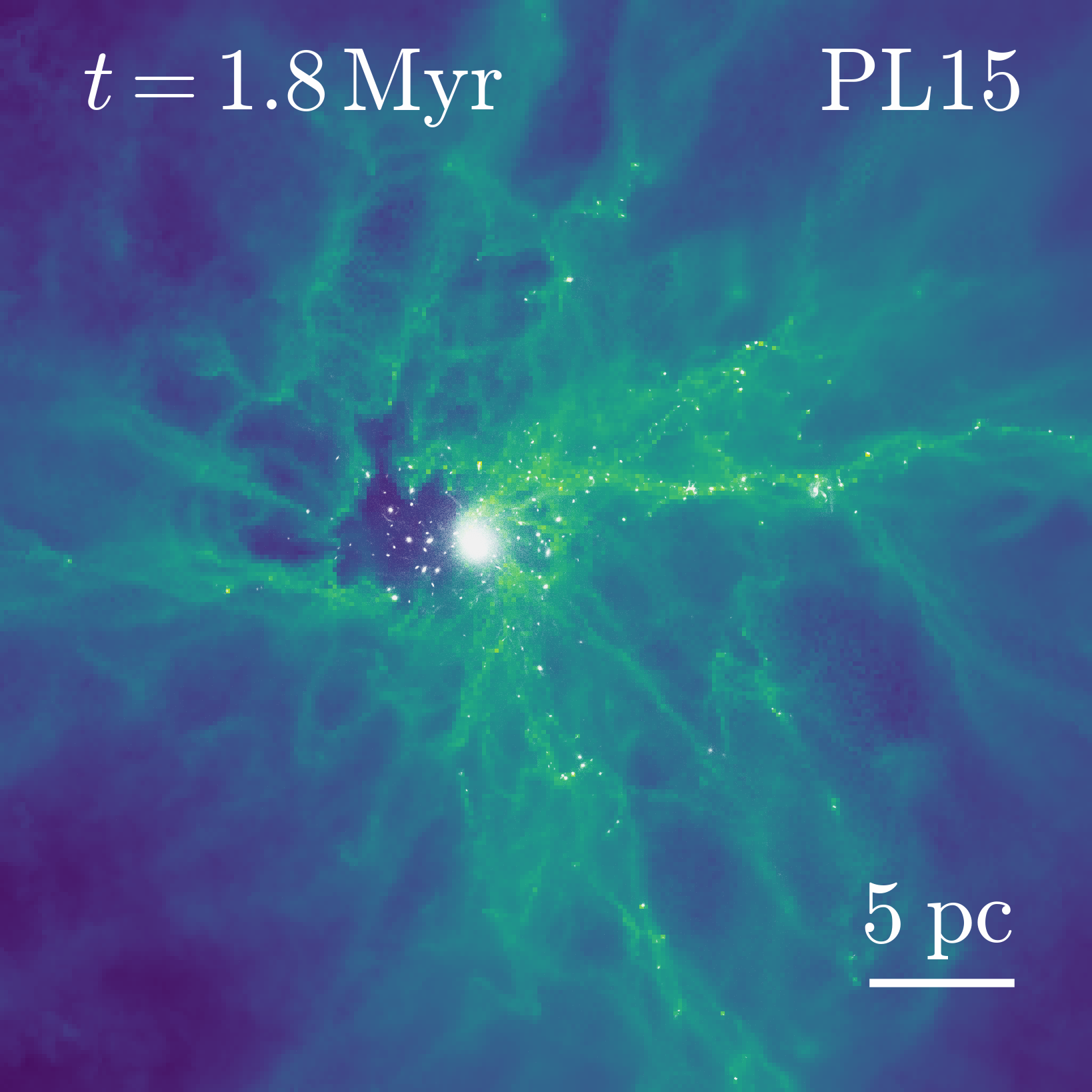}
	\vspace{0.01\columnwidth} \\
	\includegraphics[width=0.66\columnwidth]{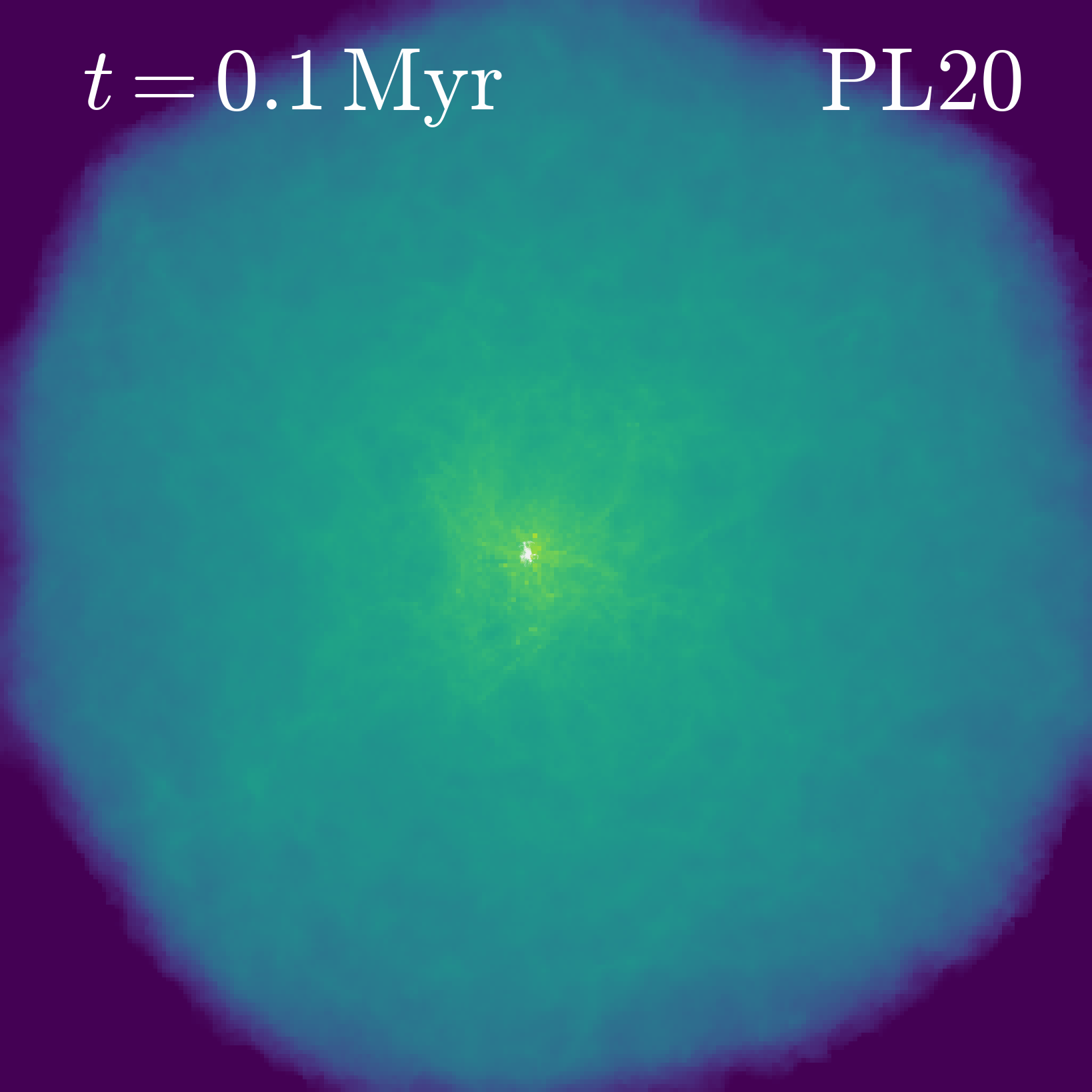}
	\hspace{0.001\columnwidth}
	\includegraphics[width=0.66\columnwidth]{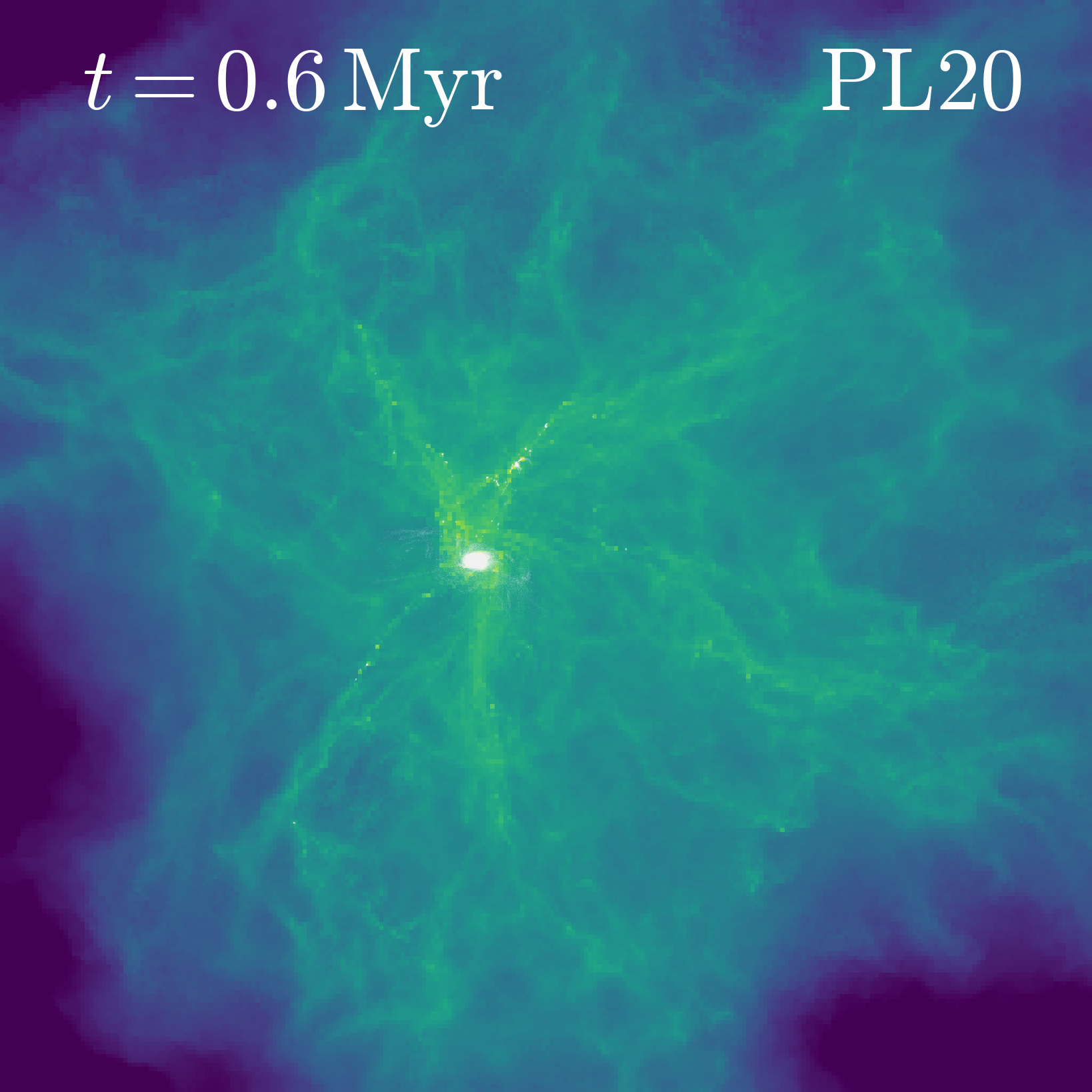}
	\hspace{0.001\columnwidth}
	\includegraphics[width=0.66\columnwidth]{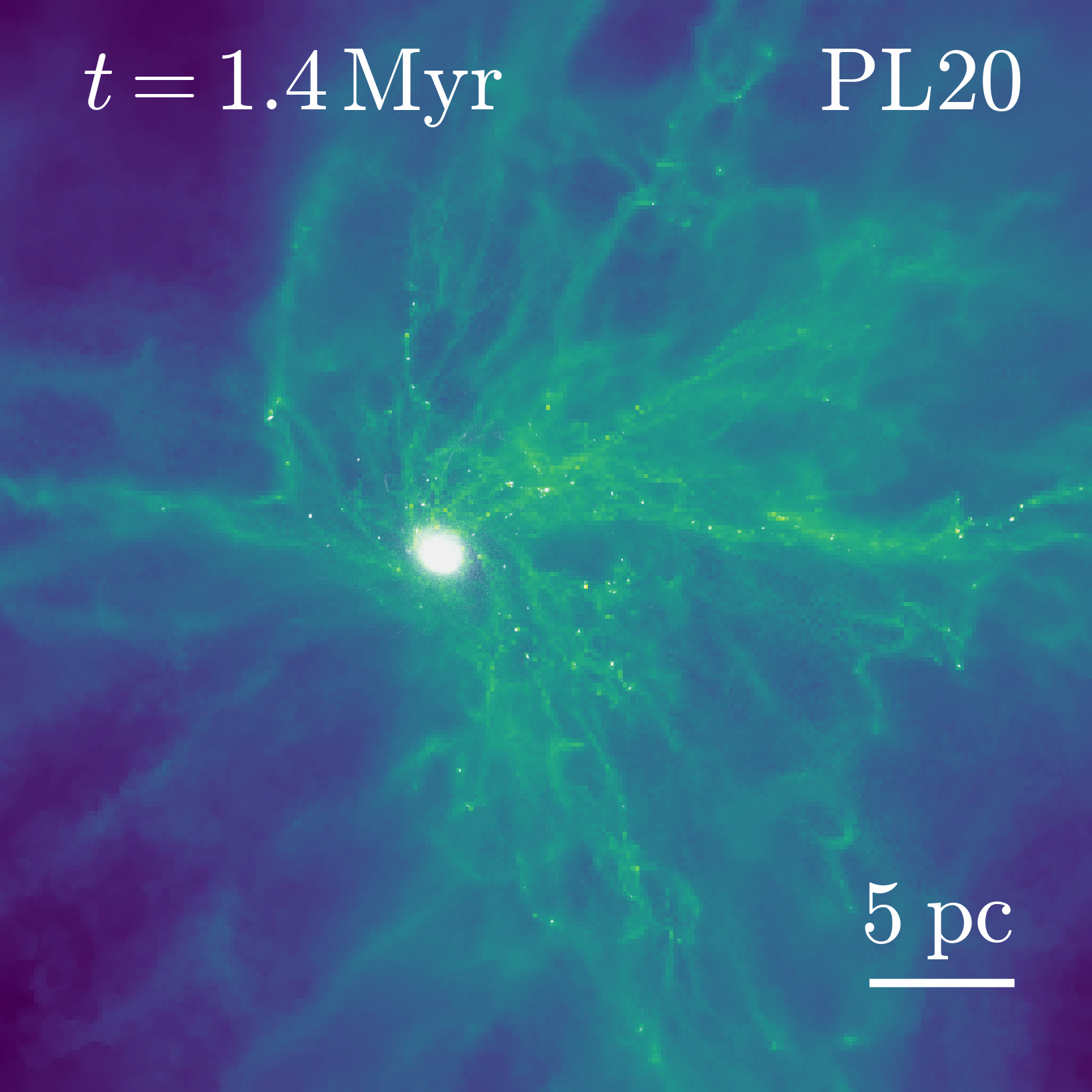}
	\vspace{0.02\columnwidth}
	\caption{Gas column density projections along the $z$-axes of the TH (top row), PL15 (middle row), and PL20 (bottom row) cases. Each plot is centered at the center of mass; stellar particles are represented as white dots. In each column, the stellar mass evolves to $10\%$ (left), $50\%$ (middle), and $90\%$ (right) of the final stellar mass, representing the initial, middle, and final stage of star formation, respectively. In each row, a length scale of $5\ \mathrm{pc}$ is shown in the lower right corner. The color range for gas column density goes from $\Sigma_{\rm gas}=10$ to $10^6\ M_{\odot}\,\mathrm{pc}^{-2}$.}
	\label{fig:prj}
\end{figure*}

\section{Results}
\label{sec:results}

First, we run each simulation with full physics as described in Sec.~\ref{sec:numericalSetup}.
All three sets of simulations with different profiles share the same general evolutionary trends.
After the dissipation of the initial turbulence velocities, gas collapses due to gravitational contraction.
Then, the cold and dense gas, mostly at the intersection between filaments, starts to fragment, and forms self-gravitating molecular cores that are gradually converted into stars.
When the stellar feedback from the central star clusters is strong enough to clear most of the gas from the central region, star formation is terminated.
Following L19, after $99\%$ of the gas mass is expelled from the initial spherical regions, we remove the remaining gas cells and continue the simulation with a pure N-body calculation to track the subsequent dynamical evolution of star clusters. 
The N-body mode is run for another couple of dynamical timescale until the most massive star cluster does not merge with any other subclusters more massive than one tenth of its mass. 

\subsection{Two modes of cluster formation: ``hierarchical'' vs. ``accretion''}
\label{sec:twoModes}

Although the evolution of the TH, PL15, and PL20 setups share a similar trend, the detailed behavior of gas assembly and star cluster formation are drastically different.
To illustrate these differences, in Fig.~\ref{fig:prj}, we show the time series of gas column density projections for runs with different initial profiles. Since the general evolution of GMCs is insensitive to random seeds, we use a representative seed as an example to demonstrate the evolution of the three initial density profiles.

For the TH setup, the gas initially forms a web-like structure with abundant filaments as a result of initial turbulence.
Many stars then form quickly within the self-gravitating molecular cores located at the intersections of these filaments.
The sub-clusters, together with some residual gas, then move along the filaments and merge into larger ones.
This merging process occurs until the clusters at the center are massive enough to exert strong stellar feedback on the nearby gas, sweeping it outwards into the low-density regions.
This star cluster formation and assembly scenario follows a  ``hierarchical'' cluster formation mode, in which a large proportion of the final central cluster's mass comes from mergers of sub-clusters that emerge from small molecular cores.

On the other hand, for the centrally-concentrated profiles, such as the PL15 and PL20 setups, the initial gas distribution shows less prominent filaments, mostly radiating from the center.
Due to the high central density, a massive star cluster is formed quickly at the central part of the cloud and continues accreting gas along the filaments until the nearby gas is expelled by the strong stellar feedback.
We call this an ``accretion'' cluster formation mode, in which the central cluster is formed early and grows its mass via accretion.

Due to different star formation modes, the TH case shows a much more spread out star formation activities across the entire cloud, while the PL15 and PL20 cases form stars mostly centrally-concentrated.
However, we stress that these two modes are not mutually exclusive; a GMC can have both ``hierarchical'' and ``accretion'' modes, yet the dominant mode determines some of the fundamental properties of star clusters, such as star formation histories, accretion rates, and stellar kinematics.

\subsection{Star formation efficiency and timescale}
\label{sec:invarianceOfSFE}

\begin{figure}
\includegraphics[width=\columnwidth]{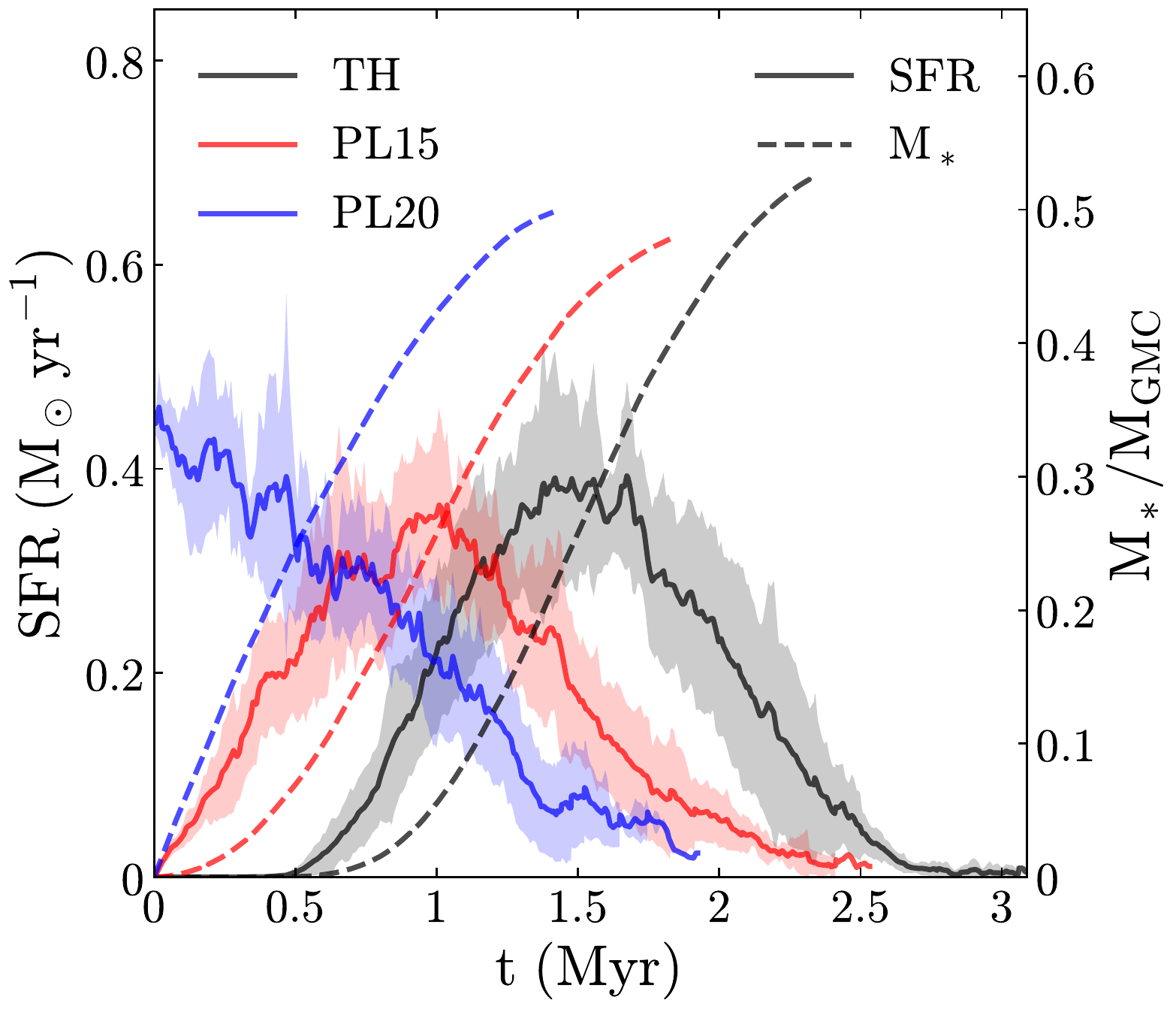}
\caption{Stellar mass (dashed) and star formation rate (solid) evolution of the TH (black), PL15 (red), and PL20 (blue) cases. The curves are calculated as mean values of ten random seeds, and the shaded areas represent the standard deviations.}
  \label{fig:SFR}
\end{figure}

In the above section, we qualitatively described the similarities and differences of the GMC evolution for the different profiles.
Now we quantify some key properties, such as star formation history, efficiency, and timescale, in detail.
We define star formation rate (SFR) as the time derivative of total stellar mass (i.e., $\mathrm{SFR}(t)=[M_*(t+\Delta t)-M_*(t)]/\Delta t$) and show the evolution of SFRs for the TH, PL15, and PL20 runs in Fig.~\ref{fig:SFR}.

The star formation history (SFH) for the TH case reproduces the findings in L19: stars are formed from $t\approx0.5\ \mathrm{Myr}\approx0.3\tau_\mathrm{half,TH}$, when a large fraction of gas collapses onto filaments.
The SFH then experiences a linear growth phase, which lasts for $0.3-0.6\tau_\mathrm{half,TH}$ before it reaches a plateau.
This linear growth phase agrees with previous analytical \citep[e.g.][]{murray_star_2015} and numerical \citep[e.g.][]{lee_time-varying_2015, murray_collapse_2017, grudic_when_2018} works.
At $t\approx \tau_\mathrm{half,TH}$, stellar feedback and the depletion of gas leads to a drop in SFR for another $\sim0.6\tau_\mathrm{half,TH}$ until the cloud is completed disrupted.
Similar to the TH scenario, the SFH of the PL15 runs also follows a linear growth phase, a plateau, and a dropping SFR stage.
These stages last for a similar amount of time to those in the TH setup.
However, the whole process occurs earlier by around $0.5\ \mathrm{Myr}$, so that the PL15 case begins star formation as soon as the simulation begins.
The SFH of the PL20 runs is distinguishable from the TH and PL15 cases in that it does not have a linear growth phase.
The initial star formation of the PL20 case is drastic, leading to a high SFR around $0.4\ M_\odot\,\mathrm{yr^{-1}}$, which is approximately the peak SFR of the TH and PL15 runs.
Additionally, for the PL20 runs, the SFR keeps decreasing after the initial star formation burst for more than $\tau_\mathrm{half,PL20}$ until it asymptotes to zero.

Surprisingly, with such drastically different modes of cluster formation and SFH, the integrated SFE and star formation duration are quite similar for the three cases.
We define the integrated SFE, $\epsilon_{\mathrm{int}}$, as the ratio of final stellar mass to the initial gas mass, $M_\mathrm{GMC}$.
As shown in Fig.~\ref{fig:SFR}, the integrated SFEs for the three profiles all have similar values of $0.501-0.542$, even with significantly different star formation histories.
To quantify the star formation timescales, we define the star formation duration as $\tau_\mathrm{dur}=t_\mathrm{90}-t_\mathrm{10}$, where $t_\mathrm{10}$ and $t_\mathrm{90}$ are the epochs when the stellar mass reaches $10\%$ and $90\%$ of the final stellar mass.
Again, the star formation duration of the three profiles have similar values of $1.04-1.17\ \mathrm{Myr}$, corresponding to $0.64-0.73\tau_\mathrm{half}$.
This indicates that the star formation duration is short and insensitive to initial profiles.
We also confirm that this conclusion is not caused by a specific choice of random seeds when generating the initial turbulent velocity fields.
We list in Tab.~\ref{tab:sum} the mean values and standard deviations of $\epsilon_{\mathrm{int}}$ and $\tau_\mathrm{dur}$ for the ten different random seeds.
The standard deviations are relatively small compared to the mean values, indicating that different seeds do not significantly alter the two properties.
Both the efficiency and timescale are largely controlled by the interplay between gravitational collapse, star formation, and stellar feedback process.

\begin{figure}
\includegraphics[width=\columnwidth]{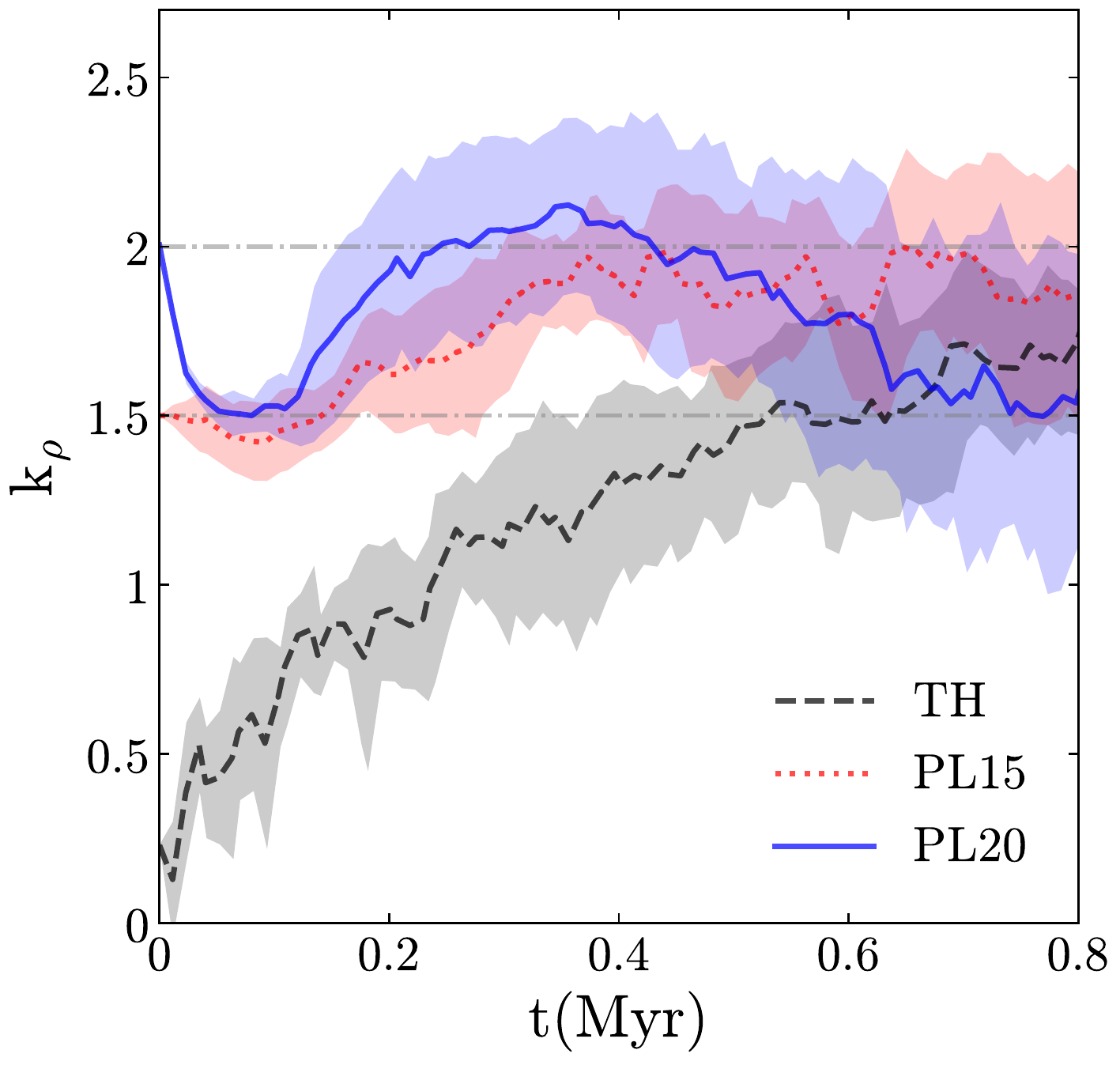}
\caption{Evolution of power-law indices, $k_\rho$, of the TH (black dashed), PL15 (red dotted), and PL20 (blue solid) cases within the central region ($0.1\ \mathrm{pc}<r<1\ \mathrm{pc}$). The two grey dash-dotted lines represent $k_\rho=1.5$ and $2$. Other plotting parameters are the same as those of Fig.~\ref{fig:SFR}.
}
  \label{fig:densProExp}
\end{figure}

\begin{figure}
\includegraphics[width=\columnwidth]{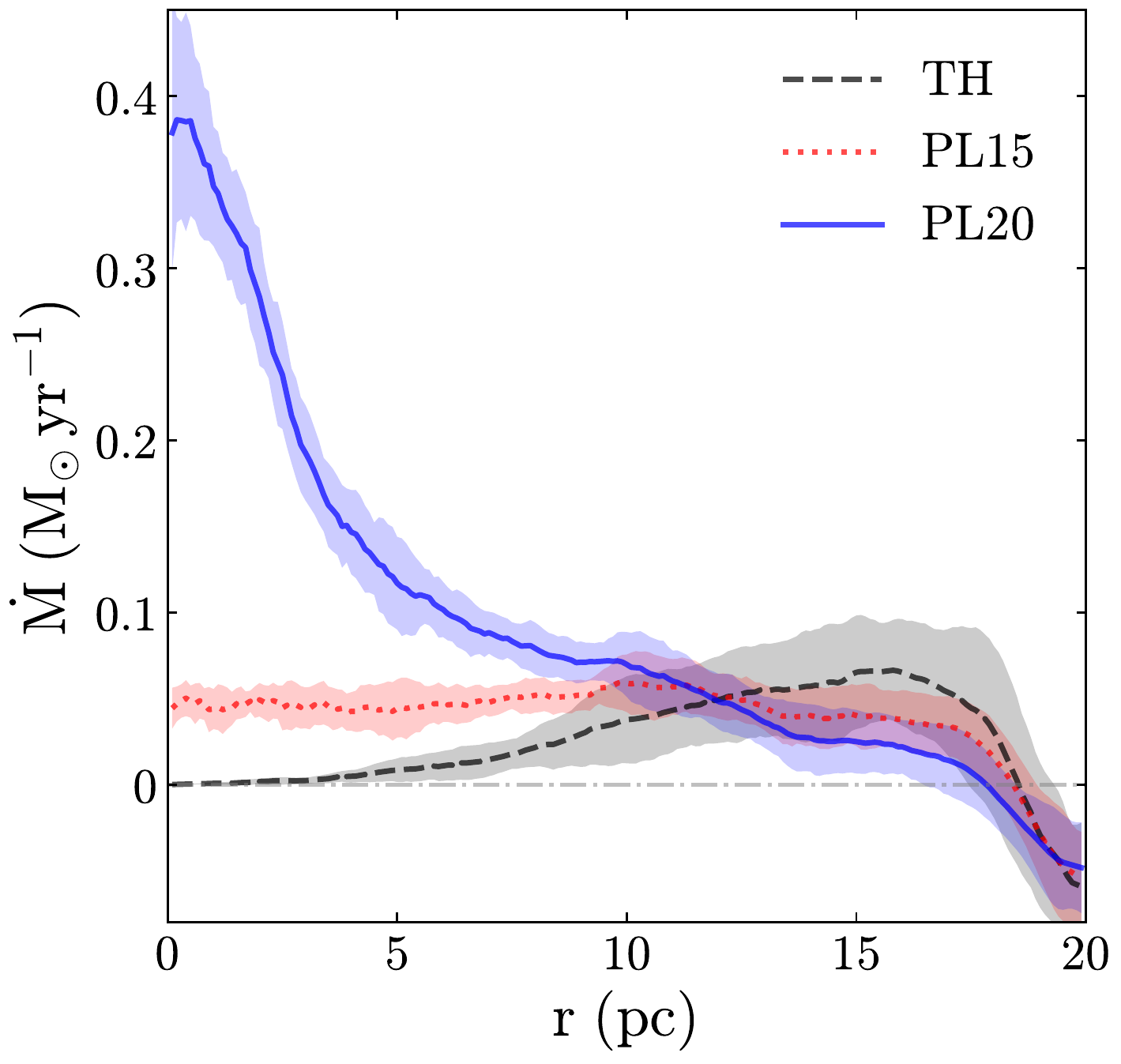}
\caption{Accretion rate profiles of the TH (black dashed), PL15 (red dotted), and PL20 (blue solid) cases at $t=0.1$. Each curve refers to the mean value of 10 random seeds. The grey dash-dotted line represents the zero accretion rate. Other plotting parameters are the same as those of Fig.~\ref{fig:SFR}.}
  \label{fig:accretionRate}
\end{figure}

\subsection{Evolution of gas density profile}
\label{sec:evolutionOfGasDensityProfile}

Previous works \citep[e.g.][]{penston_dynamics_1969, pirogov_density_2009, murray_star_2015, donkov_density_2018, li_scale-free_2018} suggested that the gas density profile of clouds follows a power-law function, $\rho(r)\propto r^{-k_\rho}$.
To investigate the evolution of the power-law index, $k_\rho$, we fit the gas density profiles with this power-law function within the central regions of GMCs, i.e., from $r=0.1$ to $1\ \mathrm{pc}$.
We plot the evolution of $k_\rho$ for the first half of the free-fall time in Fig.~\ref{fig:densProExp}.
For the TH case, $k_\rho$ rises gradually from $k_\rho=0$ to $k_\rho\approx 1.5$ at $t\approx 0.3\tau_\mathrm{half,TH}$, at the point when the TH simulations begin to form stars.
As for the PL15 runs, the $k_\rho$ curve stays nearly invariant during the initial $0.1\ \mathrm{Myr}$, then rises from $k_\rho\approx 1.5$ to $k_\rho\approx 2$ at $t\approx 0.4\ \mathrm{Myr}\approx0.2\tau_\mathrm{half,PL15}$.
Different from the above two cases, the $k_\rho$ curve of the PL20 runs falls intensely from $k_\rho\approx 2$ to $k_\rho\approx 1.5$ within the initial $0.03\ \mathrm{Myr}$, as the initial starburst (see Sec.~\ref{sec:invarianceOfSFE}) consumes the majority of the central gas. After that, it fluctuates around $k_\rho\approx 1.5$ for about $0.1\ \tau_\mathrm{half,PL20}$; this curve rises back to $k_\rho\approx 2$ again at $t\approx 0.2\ \tau_\mathrm{half,PL20}$.
All three $k_\rho$ curves behave chaotically after $t\approx0.5\ \mathrm{Myr}$, because the gas distribution in the central regions is significantly influenced by stellar feedback.

To investigate how mass flux is transferred between different radii, we analyze the gas accretion rate across the GMCs for the three initial conditions.
To obtain the accretion rate at a given radius, $r_0$, we use the following procedure.
First, we find all gas cells intersecting a sphere with radius $r_0$.
We then define the accretion rate of an intersecting cell $n$ as the multiplication of the intersecting area $A_n$, the density $\rho_n$, and the incoming velocity $u_n$, which is the compressive component of the velocity $\mathbf{v}_n$.
Finally, we calculate the accretion rate as $\Sigma_n u_n A_n \rho_n$.
In Appendix~\ref{append:accretionRate}, we compare this method with the method discussed in \citet{howard_universal_2018}.
We show that the two methods agree well with each other, with our method producing smoother curves with less fluctuations.
Using the above method, we plot the accretion rate profiles of the three cases at $t=0.1\ \mathrm{Myr}$ (i.e., at the beginning of simulations) in Fig.~\ref{fig:accretionRate}.
We note that the accretion rate of the TH runs peaks at $r\approx15\ \mathrm{pc}$ rather than the center.
As a result, the outer region of the TH case accumulates more mass (though it does not reach higher densities) than the PL15 and PL20 cases.
Therefore, the star formation activity of the TH runs is spread out over a larger region.
This is consistent with the ``hierarchical'' cluster formation mode described in Sec.~\ref{sec:twoModes}, in which the initial fragmentation in the TH runs leads to the formation of many dense molecular cores all over the GMC and is followed by the subsequent ``bottom-up'' assembly of sub-clusters toward the central cluster.
In contrast, the accretion rate profile of the PL15 runs is relatively flat, i.e., a constant mass flow from the outer region all the way to the center that keeps an invariant $\rho(r)\propto r^{-1.5}$ gas density profile.
This behavior of the PL15 case can be explained by a theoretical model proposed by \citet{murray_star_2015}, who suggested a coherent gas density profile of $\rho(r)\propto r^{-1.5}$ for a self-gravitating spherical cloud supported by turbulence.
Based on this model, we can also conclude that the PL20 profile is unable to support a prolonged star formation process as the PL15 profile.
Thus, the gas near the central region of the PL20 runs collapses quickly to center, resulting in a high initial SFR and an intense decrease of $k_\rho$ (see Fig.~\ref{fig:SFR} and \ref{fig:densProExp}).
We note that the initial accretion rate of the PL20 runs at $r\approx0$ is about $0.4\ \mathrm{M_\odot/yr}$, similar to the mass growth rate of the central cluster, indicating that the formation of central star cluster is efficiently fed by gas accretion.

\begin{figure}
\includegraphics[width=\columnwidth]{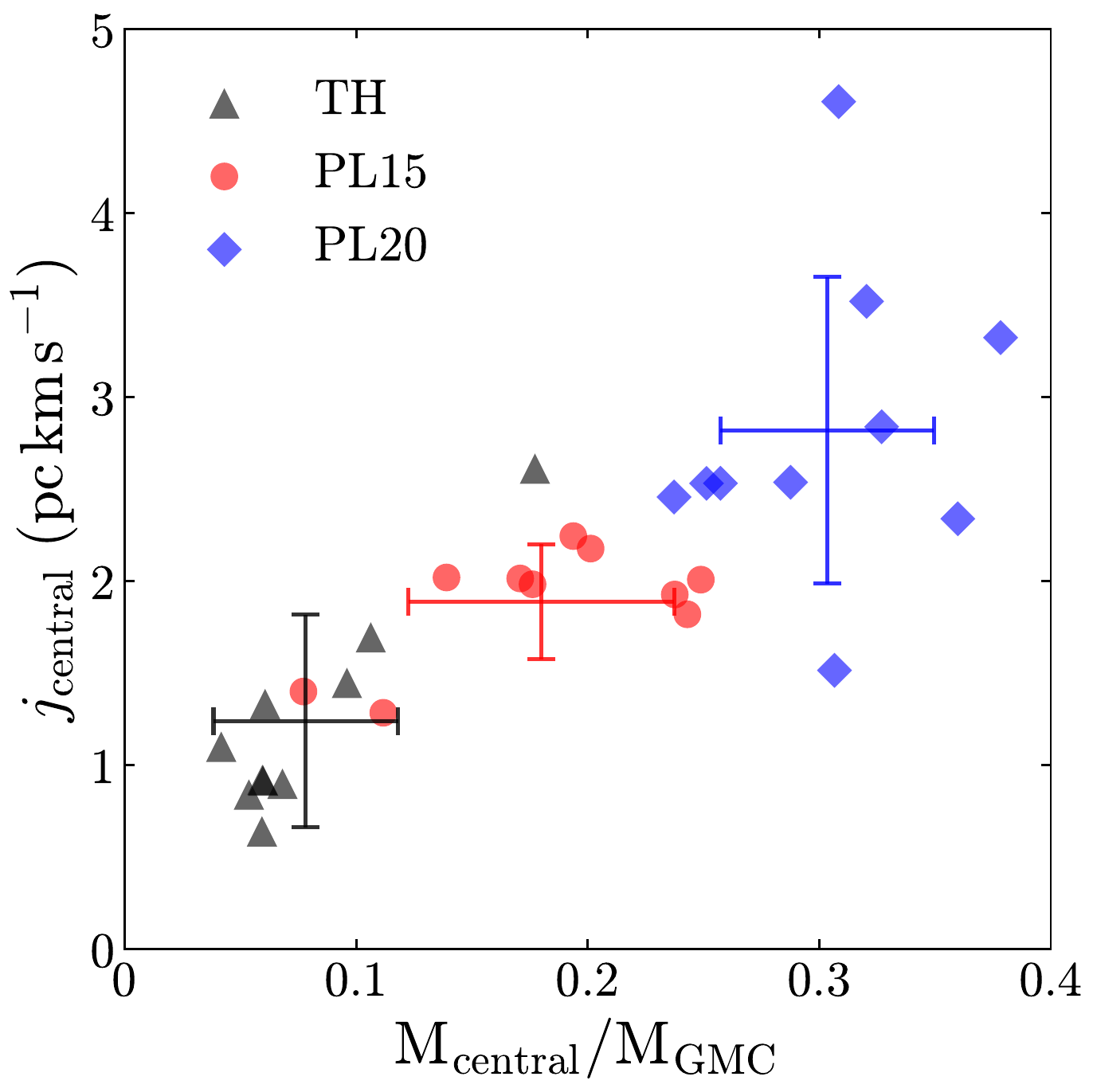}
\caption{Stellar mass -- specific angular momentum relation for central clusters in the TH (grey triangle), PL15 (red circle), and PL20 (blue diamond) cases at the final snapshots of the N-body runs. For each initial density profile, the location and size of the corresponding error-bar denote the mean values and standard deviations of $j$ and $M$, respectively.
}
  \label{fig:fdisk}
\end{figure}

\begin{figure*}
\includegraphics[width=0.66\columnwidth]{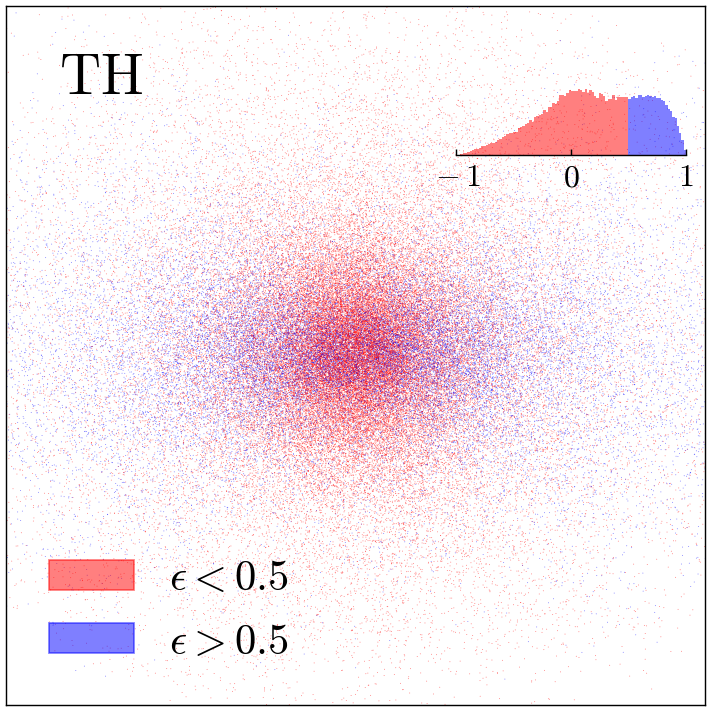}
\includegraphics[width=0.66\columnwidth]{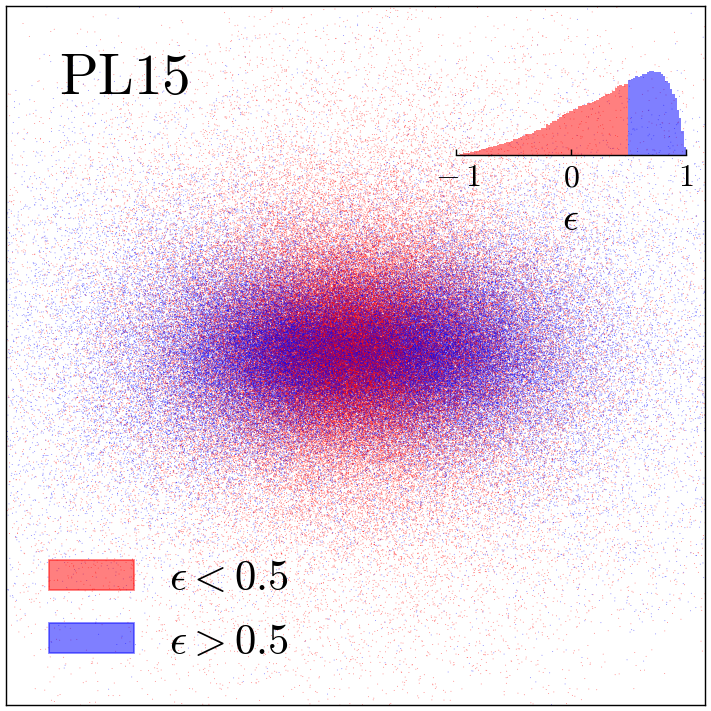}
\includegraphics[width=0.66\columnwidth]{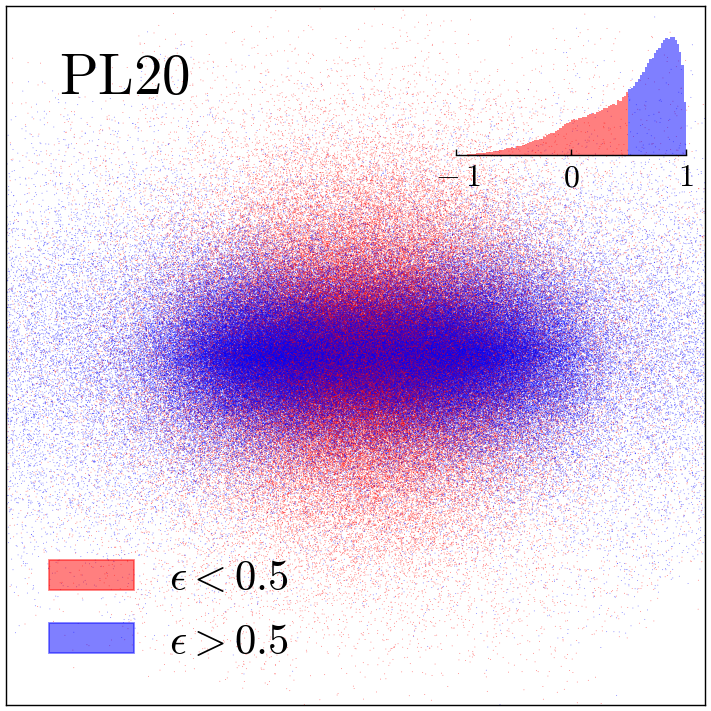}
\caption{Edge-on projections of the most massive star clusters in the TH (left), PL15 (middle), and PL20 (right) cases. Each plot is centered at the center of mass. Stellar particles are represented as blue ($\epsilon>0.5$) or red ($\epsilon<0.5$) dots. The distributions of circularity are inserted to the upper right corners.}
  \label{fig:prjCtr}
\end{figure*}

\subsection{Dynamics of substructure}
\label{sec:dynamics&Substructure}

To investigate the dynamical properties of star clusters formed from different runs, we identify gravitationally bound star clusters using the \textsc{SUBFIND} algorithm \citep{springel_populating_2001}, and calculate some key properties of the central (most massive) clusters, such as bound cluster mass ($M_\mathrm{central}$) and specific angular momentum ($j_\mathrm{central}$), at the final snapshots of the N-body simulations.
Fig.~\ref{fig:fdisk} shows the distribution of $M_\mathrm{central}/M_\mathrm{GMC}$ and $j_\mathrm{central}$ for all 30 runs.
We find that, although different random seeds produce different configurations of gas and stellar distributions, there exists a clear trend that GMCs with steeper power-law profiles form central star clusters with higher mass and specific angular momentum than those in the TH runs. To quantitatively present the above trends, we list the mean and standard deviations of $M_\mathrm{central}$ and $j_\mathrm{central}$ in Tab.~\ref{tab:sum}.
We find that $M_\mathrm{central}/M_\mathrm{GMC}$ increases from $0.078$ to $0.303$ and $j_\mathrm{central}$ from $1.24$ to $2.82\ \mathrm{pc\,km\,s^{-1}}$. We perform a one-side student-t test to compare the mean values of $M_\mathrm{central}/M_\mathrm{GMC}$ and $j_\mathrm{central}$ in different runs. We find that the p-values for all three pairs of runs (TH vs. PL15, TH vs. PL20, PL15 vs. PL20) are well below $0.05$, suggesting that both $M_\mathrm{central}/M_\mathrm{GMC}$ and $j_\mathrm{central}$ show statistically significant difference in the TH, PL15, and PL20 runs.

As suggested in Sec.~\ref{sec:twoModes}, the TH case forms star clusters hierarchically, while the centrally-concentrated cases (PL15 and PL20) form star clusters via an ``accretion'' mode.
Due to star formation activities being spread out over a much larger volume in the TH case, it is natural to expect only a small fraction of sub-clusters are end up merging into the central clusters.
This partially explains why $M_\mathrm{central}/M_\mathrm{GMC}$ is significantly smaller in clouds with shallower slopes.
Additionally, we analyze the virial parameters of the central clusters at the end of the simulations (see Tab.~\ref{tab:sum}) and find that clusters formed in clouds with steeper slopes are systematically more sub-viral than those in shallower ones.
Such a sub-virial state also helps the central clusters keep their stars during gas expulsion, see L19.

To investigate the rotation of the central star clusters, we introduce the circularity parameter, $\epsilon$, which is widely used to identify bulge and disk components in galaxy formation simulations \citep[e.g.][]{abadi_simulations_2003, martig_diversity_2012, aumer_towards_2013, marinacci_formation_2014, kannan_discs_2015, zasov_hi_2017, obreja_nihao_2016, sokolowska_galactic_2017, el-badry_gas_2018}.
Following \citet{abadi_simulations_2003, marinacci_formation_2014}, we define the circularity of a star particle $n$ in the central clusters as
\begin{equation}
    \epsilon_n=\frac{j_{z,n}}{j_c(E_n)},
    \label{eq:circularity}
\end{equation}
where $j_{z,n}$ is the angular momentum of star particle $n$ projected onto the direction of the central cluster's net angular momentum, and $j_c(E_n)$ is the maximum angular momentum for circular orbit around the cluster center for a given kinetic energy of particle $n$, $E_n$.
We call star particles with the $\epsilon>0.5$ as ``disk'' stars, and the rest as ``bulge'' stars.
Fig.~\ref{fig:prjCtr} shows the edge-on projection of central clusters as well as the distribution of circularity for different runs.
By definition, ``disk'' particles preferentially distribute along the equatorial disk of the cluster while the ``bulge'' particles form the central spheroid.
Clearly, GMCs with steeper initial profiles produce central clusters that have systematically larger ``disks''.
The mass fraction of the ``disk'' ($\epsilon>0.5$) increases from $0.379$, $0.401$ to $0.461$ for the TH, PL15, and PL20 runs, respectively.
In terms of the energy budget, central clusters in clouds with steeper profiles have higher ratios of rotational to total kinetic energy.
Interestingly, clusters formed in the PL15 and PL20 runs have $E_\mathrm{central}^\mathrm{rot}/E_\mathrm{central}^\mathrm{kin}$ larger than $20\%$, in line with the recent observations of the degree of rotation in both young massive clusters \citep[e.g.][]{henault-brunet_vlt-flames_2012} and intermediate-age clusters \citep[e.g.][]{mackey_vlt/flames_2013, kamann_linking_2019}.
In Sec.~\ref{sec:physicalOriginAM}, we will discuss the physical origins of the enhancement of specific angular momentum for GMCs with steeper profiles. 

\section{Discussion}
\label{sec:discussion}

\subsection{On the similar SFEs in different profiles}
\label{sec:theSimilarityOfSFE}

In Sec.~\ref{sec:invarianceOfSFE}, we find that the integrated SFEs for the centrally-concentrated clouds are similar to that for the TH case.
In L19, we built an analytic model that predicts the integrated SFEs for clouds with different mass, size, and feedback intensity.
In that model, stellar feedback from the central cluster sweeps nearby gas outwards, pushing/compressing it into a gas shell surrounding the cluster.
The integrated SFE is determined by the balance between gravitational contraction and the momentum injected by stellar feedback, i.e.,
\begin{equation}
\frac{M_*\dot{p}}{A_\mathrm{sh}}
=\frac{GM_*M_\mathrm{sh}}{r_\mathrm{sh}^2 A_\mathrm{sh}}
+\frac{\beta GM_\mathrm{sh}^2}{r_\mathrm{sh}^2 A_\mathrm{sh}}\:,
\label{eq:balance}
\end{equation}
where $A_\mathrm{sh}$ is the total area of the gas shell, $\beta$ is a factor quantifying the asymmetry of the gas distribution, and $\dot{p}$ is the momentum deposition rate per unit mass.
Once equilibrium is reached, any additional star formation activity will increase the momentum feedback force and disperse the gas shell.
Thus, we can calculate the integrated SFE by solving the above equation:
\begin{equation}
\epsilon_{\mathrm{int}}=\frac{M_*}{M_\mathrm{GMC}}=\frac
{\sqrt{\Gamma^2+2\,(2\beta-1)\,\Gamma+1}-(2\beta-1)\,\Gamma-1}
{2\,(1-\beta)\,\Gamma}\:,
\label{eq:solution}
\end{equation}
where $\Gamma=\pi G\Sigma_0/4\dot{p}$ such that the efficiency largely depends on the initial gas surface density, $\Sigma_0$, and feedback momentum output.
As the stellar and gas distribution in the final stage for all three cases follows a similar configuration with central clusters surrounded by gas shells, the analytical model described above is applicable to all three cases with a possible small change of $\beta$ to reflect the asymmetry of the gas shells.
Thus, we can conclude that the integrated SFE mostly correlates with the mean initial gas surface density and feedback intensity, but is insensitive to the choice of the initial density profiles.

This result is encouraging, because many previous GMC simulations which focused on predicting this efficiency do not need to worry about their specific choice of initial gas distribution.
Moreover, recent observations of nearby star-forming regions \citep[e.g.][]{zuckerman_models_1974, krumholz_slow_2007, wu_properties_2010, evans_star_2014, vutisalchavakul_star_2016, heyer_rate_2016, lee_observational_2016} reveal a large variation of the cloud-scale SFEs.
Our current work suggests that detailed structural properties of GMCs do not contribute much to the variation of the SFE.
The SFE should mainly depend on the cloud-scale properties, such as the mean gas surface density and the cloud evolutionary stage \citep{grudic_nature_2019}.
However, as discussed in Sec. \ref{sec:dynamics&Substructure}, we stress that the modes of star formation as well as various properties, such as the mass and kinematics, of the central clusters, depend strongly on the choice of initial profiles.

\subsection{On the enhancement of cluster rotation in GMCs with steeper density slopes}
\label{sec:physicalOriginAM}

\begin{figure}
	\includegraphics[width=\columnwidth]{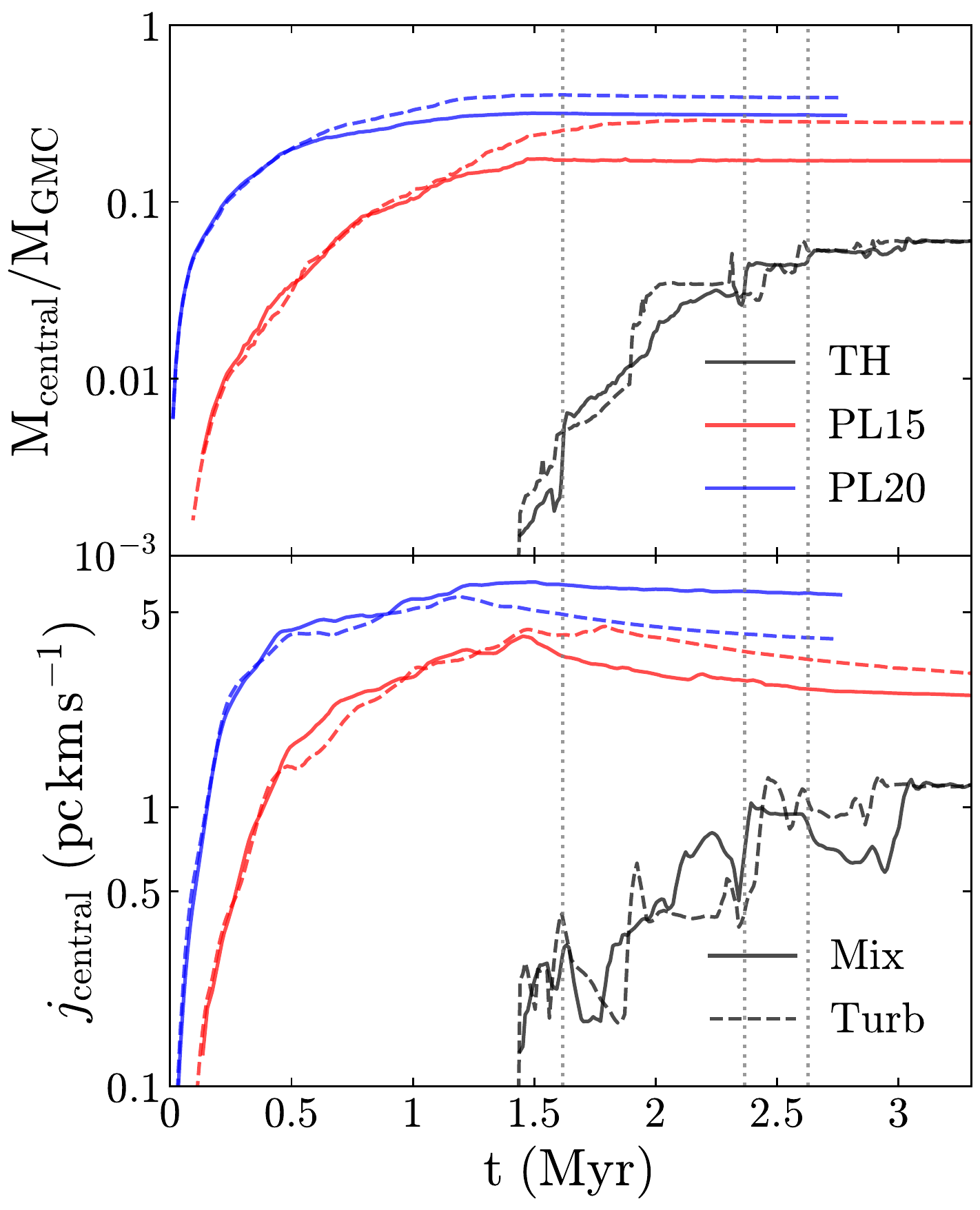}
	\caption{Mass (top) and specific angular momentum (bottom) evolution of central clusters at the final snapshots of the TH (black), PL15 (red), and PL20 (blue) cases. The solid curves represent the default ``Mix'' runs, and the dashed curves refer to the turbulence-only ``Turb'' runs. We also plot three vertical dotted lines to label the top three major mergers in the ``Mix'' run of the TH case.
	}
	\label{fig:centralFdisk}
\end{figure}

\begin{figure*}
\includegraphics[width=2\columnwidth]{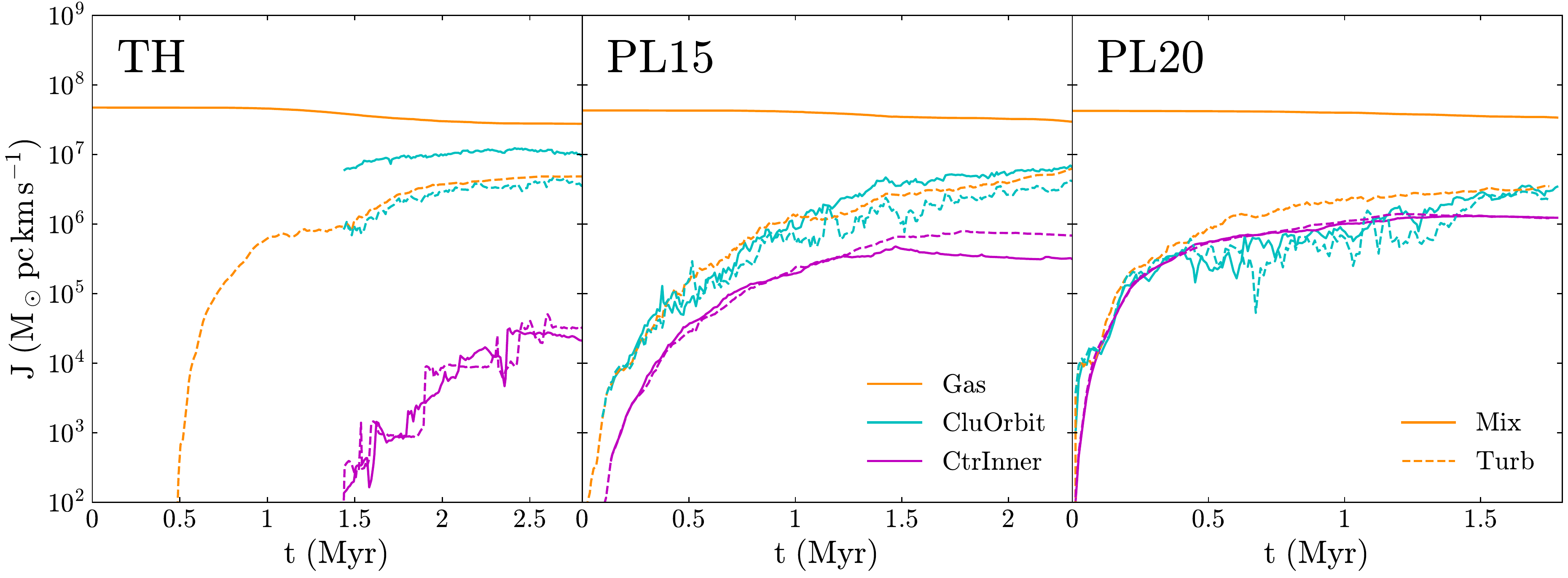}
\caption{Evolution of angular momentum from gas (Gas, orange), orbital motion of satellite clusters (CluOrbit, cyan), and inner motion of the central cluster (CtrInner, magenta). The solid curves represent the original ``Mix'' runs, and the dashed curves refer to the turbulence-only ``Turb'' runs.
}
  \label{fig:angRatio}
\end{figure*}

\begin{figure}
	\includegraphics[width=\columnwidth]{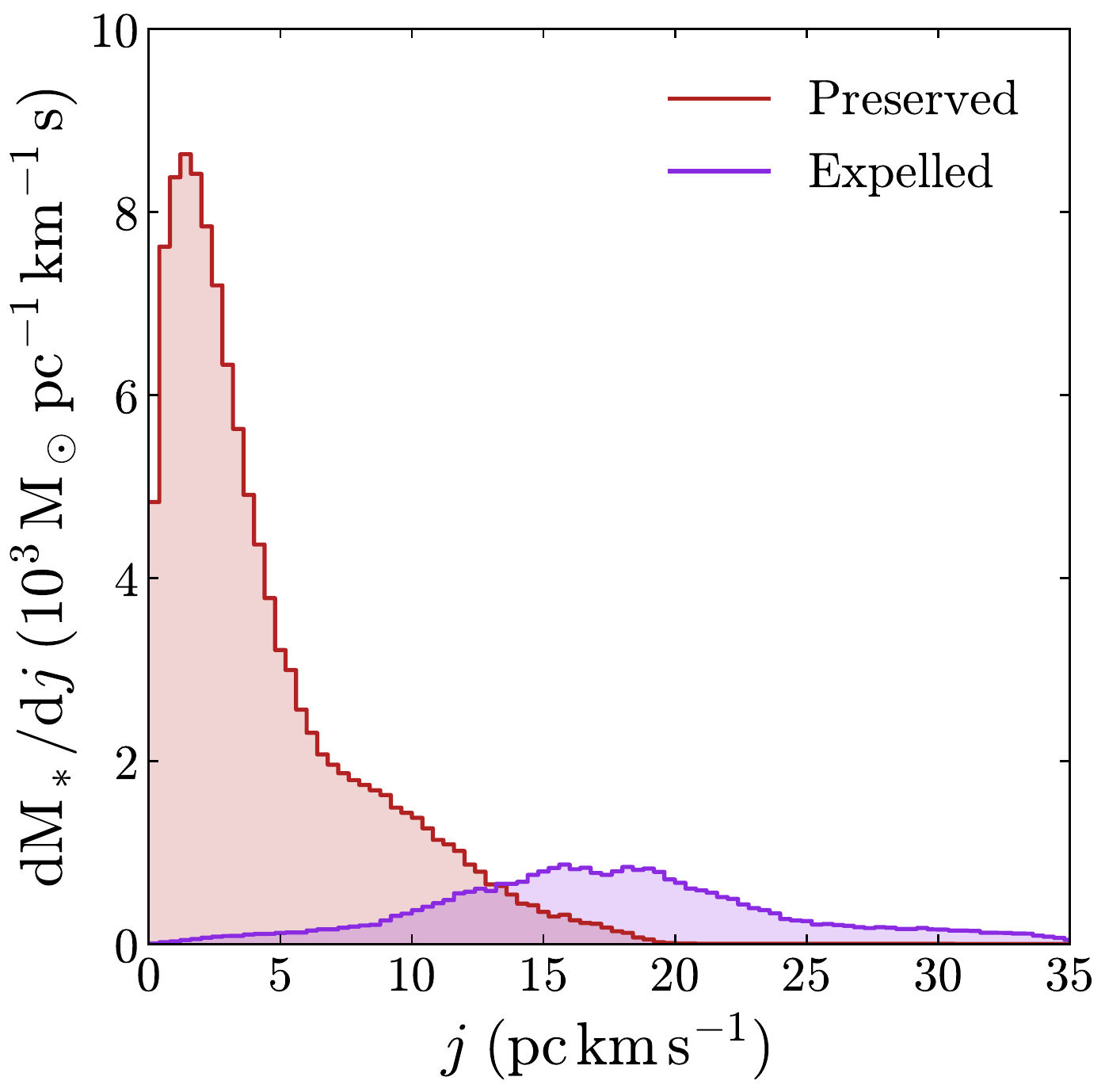}
	\caption{Distribution of specific angular momentum $0.2\ \mathrm{Myr}$ after a major merger in the N-body run of the TH case. The preserved and expelled components are colored in dark red and purple, respectively.}
	\label{fig:spinMerger}
\end{figure}

In Sec.~\ref{sec:dynamics&Substructure}, we found that the central clusters in all simulations have non-negligible angular momentum.
More importantly, we find an enhancement of the cluster rotation in more centrally-concentrated GMCs.
One natural speculation of the physical origin of this angular momentum is that it is inherited from the initial rigid rotation of the GMCs ($10\%$ of total initial kinetic energy).
To fully assess this scenario, we designed a control experiment by creating several turbulence-only rotation-free (``Turb'') counterparts of the original mixed (``Mix'') runs (see Sec.~\ref{sec:initialVelocity} for details).
We run the simulations with exactly the same physics modules and parameters as the ``Mix'' runs and compare the time evolution of the mass and specific angular momentum of the central clusters throughout their whole lifespan \footnote{We introduce a simple method to trace the evolution history of the central cluster (i.e., the main branch): starting from the last (namely, the $M$-th) snapshot, we firstly label the position, velocity, and mass of the most massive star cluster as $\mathbf{r}_M$, $\mathbf{v}_M$, and $m_M$, respectively. Then, we give prediction to the $(M-1)$-th position as $\hat{\mathbf{r}}_{M-1}=\mathbf{r}_M-\mathbf{v}_M\Delta t_M$, where $\Delta t_M$ is the duration between the two snapshots. We then find the closest $p$ star clusters to the predicted position as $p$ ``candidates''. Finally, we set the ``candidate'' with the closest mass to $m_M$ as the main progenitor at the $(M-1)$-th snapshot. We repeat the above technique over all snapshots to construct the main branch of the star cluster evolution. We test the robustness of the method by changing the value of $p$ from $2$ to $16$, and find it does not affect result.}.
In Fig.~\ref{fig:centralFdisk}, we show the mass and specific angular momentum evolution of the main branches from both the ``Mix'' and ``Turb'' runs. We find that the absence of initial angular momentum in the ``Turb'' runs does not significantly affect the evolution of both the mass and angular momentum of central clusters.
In some cases, $j_{\rm central}$ in the ``Turb'' runs is even slightly larger than that in ``Mix'' runs. This indicates that the angular momentum of central clusters \textit{does not} directly stem from the initial rigid rotation of the GMCs, but originates from the symmetry-break of the turbulent velocity field.

During the course of cluster evolution, $j_{\rm central}$ is always higher for runs with steeper density slopes.
To better understand this trend, we separate the total angular momentum into three different components: 1) the angular momentum of gas (Gas); 2) the orbital angular momentum of sub-clusters around the central cluster (CluOrbit); 3) the intrinsic rotation of the central star cluster (CtrInner).
Fig.~\ref{fig:angRatio} illustrates the evolution of the three components of the TH, PL15, and PL20 runs.
First, the angular momentum stored in gas dominates the total angular momentum budget of the GMC-cluster system and is similar for all three cases.
Second, although the total stellar angular momentum (CluOrbit+CtrInner) for all three cases are similar, almost all stellar angular momentum of TH is stored in the orbital motion of sub-clusters rather than the rotation of central star clusters.
In contrast, the contribution of the two components is similar in the PL15 and PL20 cases.
This means that the reason why central clusters emerged from the TH runs show much less rotation than that from the PL15 and PL20 runs is that a large number of sub-clusters, which store the dominant angular momentum, are not able to merge into the central cluster due to the scattered stellar distribution.

Additionally, for sub-clusters that eventually emerged into the central cluster in the TH runs, these merger events inevitably reduce the specific angular momentum of the system via violent relaxation and tidal disruption.
In Fig.~\ref{fig:centralFdisk}, we find that the mass and (specific) angular momentum evolution of the TH case undergoes more frequent sudden rises than those of the PL15 and PL20 runs.
Such discontinuities reflect the epochs of major mergers events in the TH case.
To investigate how major mergers influence the mass and angular momentum of central clusters, we focus on one specific major merger event in the N-body run of the TH case.
We plot the distribution of specific angular momentum of the stellar particles that are still bound to the central cluster and of those that are expelled $0.2\ \mathrm{Myr}$ after the merger in Fig.~\ref{fig:spinMerger}. We did the same analysis $0.1-0.4\ \mathrm{Myr}$ after the merger and found that the distribution does not change.
The escaping stellar particles show a broad range of specific angular momentum with a peak at $j\sim15-20$ pc km/s, significantly larger than the peak of the preserved ones. This is expected since stellar particles with larger specific angular momentum are more likely to escape during merger.
As we have already seen in Fig.~\ref{fig:centralFdisk}, central clusters in TH runs are assebled in a ``hierarachical'' fashion and experience frequent major mergers, each of which gets rid of a fraction of the angular momentum of the system.
In contrast, for clusters formed in PL15 and PL20 runs, gas is accreted directly to the central cluster before forming sub-clusters, so the mass and angular momentum are more efficiently transferred to the central cluster in early stages.
This is another physical reason why clusters emerged from steeper density slopes rotate faster.

\subsection{Rotation of most massive star clusters}
\label{sec:rotationOfClusters}

Globular clusters (GCs) have long been considered as well-relaxed non-rotating systems.
However, recent observations have revealed a large number of Galactic GCs with statistically significant rotation \citep[e.g.][]{bellazzini_na-o_2012, lardo_gaia_2015, bianchini_internal_2018, kamann_stellar_2018, ferraro_mikis_2018}.
More interestingly, among all rotating GCs, there exists a positive correlation between the rotation and relaxation time (or mass) of the clusters \citep{bianchini_internal_2018, kamann_stellar_2018}, suggesting that the rotation in GCs is likely the relic of the internal rotation when the clusters emerged from their natal GMCs at high-z.
This scenario is strengthened by recent observations that revealed rotation signatures in both young \citep{fischer_dynamics_1992, fischer_dynamics_1993, henault-brunet_vlt-flames_2012} and intermediate-age \citep{mackey_vlt/flames_2013, kamann_linking_2019} star clusters. Essentially, internal rotation has become a common dynamical configuration for star clusters of all ages.

Our simulations propose a promising mechanism of forming highly rotating clusters from the ``accretion'' mode of cluster formation in centrally-concentrated clouds.
The ``accretion'' mode (e.g., in PL20 runs) helps gas channel its angular momentum all the way to the center of the cloud before it fragments to form stars.
Therefore, the central cluster receives angular momentum from a large volume of gas reservoir that contains large amount of angular momentum.
This scenario is in line with the most recent galaxy formation simulations that resolve the formation of individual star clusters in high-pressure turbulent disks at high-redshifts \citep{ma_self-consistent_2020} and mergers of gas-rich dwarf galaxies \citep{lahen_griffin_2020}.
The strong rotation of young massive clusters also strongly affect their long-term dynamical evolution. It has been suggested that internal rotation accelerates the core collapse via gravogyro instability and increase the mass loss rate \cite[e.g.][]{einsel_dynamical_1999, ernst_n_2007, tiongco_complex_2018}. Models that focus on the dynamical evolution of GCs, especially some recent efforts that track GC evolution in realistic galactic environments \citep[e.g.][]{li_star_2017, pfeffer_e-mosaics_2018, li_star_2019}, should include the effect of cluster rotation to better evaluate the mass-loss rate of GCs across cosmic time.

\subsection{Comparison to previous simulations}
\label{sec:comparison}

Over the past decade, there have been a few numerical studies that investigated GMCs from similar perspectives as our work. We now discuss how these studies relate to the work presented here.

\citet{girichidis_importance_2011, girichidis_importance_2012, girichidis_importance_2012-1} performed a suite of simulations of dense molecular cores with different initial conditions: two flat initial density profiles (TH and Bonnor-Ebert sphere) and two centrally-concentrated power-law profiles (PL15 and PL20).
Although using very different numerical setup and different mass and size of the clouds, some of their results are similar to ours.
They found that the PL15 and PL20 runs show much earlier star formation activities than the TH run, consistent with our result. Moreover, they found that the flat density profiles produce many low-mass stars distributed throughout the entire cloud, whileS concentrated profiles only form one massive particle at the center. This trend is in line with our ``hierarchical'' and ``accretion'' cluster formation modes.

In fact, the ``hierarchical'' nature of cluster formation is commonly seen in previous GMC simulations. For example, in \citet{grudic_top_2018}, they described cluster formation as a combination of the ``top-down'' fragmentation and the subsequent ``bottom-up'' assembly of sub-clusters. \citet{howard_universal_2018} quantitatively analyzed the hierarchical nature of cluster formation and suggested about $50\%$ of the central cluster's mass comes from mergers. It should be noted, however, that most of their conclusions are based on simulations with flat density profiles. As we show in this work, for clouds with steeper density profiles, both the ``top-down'' fragmentation and ``bottom-up'' sub-cluster assembly becomes rarer, leading to a significant deviation from the ``hierarchical'' cluster formation mode.

Another important result of our work is the rotation of central clusters. Previous works, such as \citet[][]{mapelli_rotation_2017, ballone_evolution_2020}, found rotation signatures of star clusters in their simulations. They found that the angular momentum of star clusters is acquired from their parent gas due to large-scale torques during the the process of hierarchical assembly. This is consistent with our TH runs where central clusters contain $\sim18\%$ of total kinetic energy in rotation. However, we stress that ``hirarchical'' assembly of star clusters is not the main driver of cluster rotation. In contrast, clusters formed in ``accretion'' mode tend to be more rotational-supported.

\section{Summary}
\label{sec:summary}

In this paper, we present a suite of hydrodynamic simulations of GMCs, employing the quasi-Lagrangian moving-mesh code \textsc{Arepo} with explicit cooling, star formation, and stellar feedback models. We investigate star cluster formation in GMCs with three different initial gas density profiles: $\rho(r)= r^0$, $\rho(r)\propto r^{-1.5}$, and $\rho(r)\propto r^{-2}$. 
Our main conclusions are as follows:

\begin{itemize}
    \item GMCs with different initial density profiles show drastically different modes of star cluster formation. Clouds with shallower profiles, such as TH, create significant filamentary structures across the entire region.
    The filaments then fragment and form abundant self-gravitating molecular cores, where many sub-clusters are formed subsequently. Some of the sub-clusters gradually merge into central clusters, reflecting a ``hierarchical'' cluster formation mode. In contrast, clouds with centrally-concentrated profiles, such as PL15 and PL20, quickly form one massive star cluster in the center of the cloud and then start accreting gas from the ambient region. The accretion rate towards the center is high so that gas does not have enough time to fragment and form stars before it is fully compressed at the center-most region. This ``accretion'' cluster formation mode creates highly compact central clusters.
    
    \item Although experiencing different cluster formation modes, the overall star formation properties, such as the star formation duration and integrated SFE, are very similar for clouds with different initial profiles. Both quantities are controlled by the interplay between gravitational collapse and stellar feedback. Following the analytical model in L19, the integrated SFE mostly correlates with the mean initial surface density and stellar feedback intensity, but weakly depends on the choice of initial density profiles.
    
    \item The accretion rate profile of the TH case peaks at $r\approx 15\ \mathrm{pc}$, resulting in a more extended spatial distribution of star formation than the PL15 and PL20 cases. Because of the flat accretion rate profile, the PL15 profile can maintain a stabilized gas density profile as well as a linear rising SFR before stellar feedback disrupts the clouds. For the PL20 case, the accretion rate is initially high enough to support an intense star formation, leading to a starburst at the center of the cloud in the early evolution stage.
    
    \item Clouds with steeper profiles produce more massive central clusters with higher specific angular momentum than the shallower profile cases. This is because 1) a larger proportion of mass and angular momentum in the shallower cases is stored in the orbiting sub-clusters that are not able to merge into the central clusters; 2) frequent major mergers in the shallower profiles lead to further losses of mass and angular momentum via violent relaxation and tidal disruption. In contract, clouds with steeper profiles transfer mass and angular momentum to the most massive clusters mostly via smooth accretion, which better preserves angular momentum.
    
    \item We find that rotation is a common kinematic signature in model clusters. Encouragingly, the degree of cluster rotations in PL15 and PL20 runs is consistent with recent observations of young and intermediate-age clusters. We speculate that rotating globular clusters are likely formed via an ``accretion'' mode from centrally-concentrated clouds in the early Universe. Models that focus on dynamical evolution of GCs should include the effect of cluster rotation to better evaluate the mass-loss rate of GCs across cosmic time.
\end{itemize}

\section*{Acknowledgements}
We thank Volker Springel for giving us access to AREPO. We are grateful to David Barnes, Laura Sales, Eve Ostriker, Oleg Gnedin, and Vadim Semenov for insightful comments and suggestions. H.L. is supported by NASA through the NASA Hubble Fellowship grant HST-HF2-51438.001-A awarded by the Space Telescope Science Institute, which is operated by the Association of Universities for Research in Astronomy, Incorporated, under NASA contract NAS5-26555. MV acknowledges support through an MIT RSC award, a Kavli Research Investment Fund, NASA ATP grant NNX17AG29G, and NSF grants AST-1814053, AST-1814259 and AST-1909831. The simulations of this work were run on the Harvard Odyssey clusters and the Comet HPC resource at San Diego Supercomputer Center as part of XSEDE through TG-AST170042 and TG-AST180025.




\bibliographystyle{mnras}
\bibliography{GMC-profiles-reference} 

\begin{thebibliography}{}
\makeatletter
\relax
\def\mn@urlcharsother{\let\do\@makeother \do\$\do\&\do\#\do\^\do\_\do\%\do\~}
\def\mn@doi{\begingroup\mn@urlcharsother \@ifnextchar [ {\mn@doi@}
  {\mn@doi@[]}}
\def\mn@doi@[#1]#2{\def\@tempa{#1}\ifx\@tempa\@empty \href
  {http://dx.doi.org/#2} {doi:#2}\else \href {http://dx.doi.org/#2} {#1}\fi
  \endgroup}
\def\mn@eprint#1#2{\mn@eprint@#1:#2::\@nil}
\def\mn@eprint@arXiv#1{\href {http://arxiv.org/abs/#1} {{\tt arXiv:#1}}}
\def\mn@eprint@dblp#1{\href {http://dblp.uni-trier.de/rec/bibtex/#1.xml}
  {dblp:#1}}
\def\mn@eprint@#1:#2:#3:#4\@nil{\def\@tempa {#1}\def\@tempb {#2}\def\@tempc
  {#3}\ifx \@tempc \@empty \let \@tempc \@tempb \let \@tempb \@tempa \fi \ifx
  \@tempb \@empty \def\@tempb {arXiv}\fi \@ifundefined
  {mn@eprint@\@tempb}{\@tempb:\@tempc}{\expandafter \expandafter \csname
  mn@eprint@\@tempb\endcsname \expandafter{\@tempc}}}

\bibitem[\protect\citeauthoryear{Abadi, Navarro, Steinmetz  \& Eke}{Abadi
  et~al.}{2003}]{abadi_simulations_2003}
Abadi M.~G.,  Navarro J.~F.,  Steinmetz M.,   Eke V.~R.,  2003, \mn@doi [ApJ]
  {10.1086/378316}, 597, 21

\bibitem[\protect\citeauthoryear{Aumer, White, Naab  \& Scannapieco}{Aumer
  et~al.}{2013}]{aumer_towards_2013}
Aumer M.,  White S. D.~M.,  Naab T.,   Scannapieco C.,  2013, \mn@doi [MNRAS]
  {10.1093/mnras/stt1230}, 434, 3142

\bibitem[\protect\citeauthoryear{Ballone, Mapelli, Di~Carlo, Torniamenti, Spera
   \& Rastello}{Ballone et~al.}{2020}]{ballone_evolution_2020}
Ballone A.,  Mapelli M.,  Di~Carlo U.~N.,  Torniamenti S.,  Spera M.,
  Rastello S.,  2020, arXiv:2001.10003 [astro-ph]

\bibitem[\protect\citeauthoryear{Barnett, Magland  \& af Klinteberg}{Barnett
  et~al.}{2018}]{barnett_parallel_2018}
Barnett A.~H.,  Magland J.~F.,   af Klinteberg L.,  2018, arXiv:1808.06736 [cs,
  math]

\bibitem[\protect\citeauthoryear{Bellazzini, Bragaglia, Carretta, Gratton,
  Lucatello, Catanzaro  \& Leone}{Bellazzini
  et~al.}{2012}]{bellazzini_na-o_2012}
Bellazzini M.,  Bragaglia A.,  Carretta E.,  Gratton R.~G.,  Lucatello S.,
  Catanzaro G.,   Leone F.,  2012, \mn@doi [A\&A]
  {10.1051/0004-6361/201118056}, 538, A18

\bibitem[\protect\citeauthoryear{Bianchini, {van der Marel}, {del Pino},
  Watkins, Bellini, Fardal, Libralato  \& Sills}{Bianchini
  et~al.}{2018}]{bianchini_internal_2018}
Bianchini P.,  {van der Marel} R.~P.,  {del Pino} A.,  Watkins L.~L.,  Bellini
  A.,  Fardal M.~A.,  Libralato M.,   Sills A.,  2018, \mn@doi [MNRAS]
  {10.1093/mnras/sty2365}, 481, 2125

\bibitem[\protect\citeauthoryear{Bonnell, Bate  \& Vine}{Bonnell
  et~al.}{2003}]{bonnell_hierarchical_2003}
Bonnell I.~A.,  Bate M.~R.,   Vine S.~G.,  2003, \mn@doi [MNRAS]
  {10.1046/j.1365-8711.2003.06687.x}, 343, 413

\bibitem[\protect\citeauthoryear{Burkert \& Hartmann}{Burkert \&
  Hartmann}{2013}]{burkert_dependence_2013}
Burkert A.,  Hartmann L.,  2013, \mn@doi [ApJ] {10.1088/0004-637X/773/1/48},
  773, 48

\bibitem[\protect\citeauthoryear{Csengeri et~al.,}{Csengeri
  et~al.}{2017}]{csengeri_alma_2017}
Csengeri T.,  et~al., 2017, \mn@doi [A\&A] {10.1051/0004-6361/201629754}, 600,
  L10

\bibitem[\protect\citeauthoryear{Dale, Ercolano  \& Bonnell}{Dale
  et~al.}{2012}]{dale_ionizing_2012}
Dale J.~E.,  Ercolano B.,   Bonnell I.~A.,  2012, \mn@doi [MNRAS]
  {10.1111/j.1365-2966.2012.21205.x}, 424, 377

\bibitem[\protect\citeauthoryear{Dale, Ngoumou, Ercolano  \& Bonnell}{Dale
  et~al.}{2014}]{dale_before_2014}
Dale J.~E.,  Ngoumou J.,  Ercolano B.,   Bonnell I.~A.,  2014, \mn@doi [MNRAS]
  {10.1093/mnras/stu816}, 442, 694

\bibitem[\protect\citeauthoryear{Dobbs et~al.,}{Dobbs
  et~al.}{2014}]{dobbs_formation_2014}
Dobbs C.~L.,  et~al., 2014, in , Protostars and {{Planets VI}}.
{University of Arizona Press},
  \mn@doi{10.2458/azu_uapress_9780816531240-ch001}, \url
  {http://muse.jhu.edu/books/9780816598762/9780816598762-7.pdf}

\bibitem[\protect\citeauthoryear{Donkov \& Stefanov}{Donkov \&
  Stefanov}{2018}]{donkov_density_2018}
Donkov S.,  Stefanov I.~Z.,  2018, \mn@doi [MNRAS] {10.1093/mnras/stx3116},
  474, 5588

\bibitem[\protect\citeauthoryear{Einsel \& Spurzem}{Einsel \&
  Spurzem}{1999}]{einsel_dynamical_1999}
Einsel C.,  Spurzem R.,  1999, \mn@doi [MNRAS]
  {10.1046/j.1365-8711.1999.02083.x}, 302, 81

\bibitem[\protect\citeauthoryear{{El-Badry} et~al.,}{{El-Badry}
  et~al.}{2018}]{el-badry_gas_2018}
{El-Badry} K.,  et~al., 2018, \mn@doi [MNRAS] {10.1093/mnras/stx2482}, 473,
  1930

\bibitem[\protect\citeauthoryear{Ernst, Glaschke, Fiestas, Just  \&
  Spurzem}{Ernst et~al.}{2007}]{ernst_n_2007}
Ernst A.,  Glaschke P.,  Fiestas J.,  Just A.,   Spurzem R.,  2007, \mn@doi
  [MNRAS] {10.1111/j.1365-2966.2007.11602.x}, 377, 465

\bibitem[\protect\citeauthoryear{Evans, Heiderman  \& Vutisalchavakul}{Evans
  et~al.}{2014}]{evans_star_2014}
Evans N.~J.,  Heiderman A.,   Vutisalchavakul N.,  2014, \mn@doi [ApJ]
  {10.1088/0004-637X/782/2/114}, 782, 114

\bibitem[\protect\citeauthoryear{Fall, Krumholz  \& Matzner}{Fall
  et~al.}{2010}]{fall_stellar_2010}
Fall S.~M.,  Krumholz M.~R.,   Matzner C.~D.,  2010, \mn@doi [ApJ]
  {10.1088/2041-8205/710/2/L142}, 710, L142

\bibitem[\protect\citeauthoryear{Ferraro et~al.,}{Ferraro
  et~al.}{2018}]{ferraro_mikis_2018}
Ferraro F.~R.,  et~al., 2018, \mn@doi [ApJ] {10.3847/1538-4357/aabe2f}, 860, 50

\bibitem[\protect\citeauthoryear{Fischer, Welch, Cote, Mateo  \&
  Madore}{Fischer et~al.}{1992}]{fischer_dynamics_1992}
Fischer P.,  Welch D.~L.,  Cote P.,  Mateo M.,   Madore B.~F.,  1992, \mn@doi
  [AJ] {10.1086/116107}, 103, 857

\bibitem[\protect\citeauthoryear{Fischer, Welch  \& Mateo}{Fischer
  et~al.}{1993}]{fischer_dynamics_1993}
Fischer P.,  Welch D.~L.,   Mateo M.,  1993, \mn@doi [AJ] {10.1086/116483},
  105, 938

\bibitem[\protect\citeauthoryear{Geen, Hennebelle, Tremblin  \& Rosdahl}{Geen
  et~al.}{2015}]{geen_photoionization_2015}
Geen S.,  Hennebelle P.,  Tremblin P.,   Rosdahl J.,  2015, \mn@doi [MNRAS]
  {10.1093/mnras/stv2272}, 454, 4484

\bibitem[\protect\citeauthoryear{Girichidis, Federrath, Banerjee  \&
  Klessen}{Girichidis et~al.}{2011}]{girichidis_importance_2011}
Girichidis P.,  Federrath C.,  Banerjee R.,   Klessen R.~S.,  2011, \mn@doi
  [MNRAS] {10.1111/j.1365-2966.2011.18348.x}, 413, 2741

\bibitem[\protect\citeauthoryear{Girichidis, Federrath, Banerjee  \&
  Klessen}{Girichidis et~al.}{2012a}]{girichidis_importance_2012}
Girichidis P.,  Federrath C.,  Banerjee R.,   Klessen R.~S.,  2012a, \mn@doi
  [MNRAS] {10.1111/j.1365-2966.2011.20073.x}, 420, 613

\bibitem[\protect\citeauthoryear{Girichidis, Federrath, Allison, Banerjee  \&
  Klessen}{Girichidis et~al.}{2012b}]{girichidis_importance_2012-1}
Girichidis P.,  Federrath C.,  Allison R.,  Banerjee R.,   Klessen R.~S.,
  2012b, \mn@doi [MNRAS] {10.1111/j.1365-2966.2011.20250.x}, 420, 3264

\bibitem[\protect\citeauthoryear{Girichidis et~al.,}{Girichidis
  et~al.}{2020}]{girichidis_physical_2020}
Girichidis P.,  et~al., 2020, arXiv:2005.06472 [astro-ph.GA]

\bibitem[\protect\citeauthoryear{Grudi{\'c}, Hopkins, {Faucher-Gigu{\`e}re},
  Quataert, Murray  \& Kere{\v s}}{Grudi{\'c} et~al.}{2018a}]{grudic_when_2018}
Grudi{\'c} M.~Y.,  Hopkins P.~F.,  {Faucher-Gigu{\`e}re} C.-A.,  Quataert E.,
  Murray N.,   Kere{\v s} D.,  2018a, \mn@doi [MNRAS] {10.1093/mnras/sty035},
  475, 3511

\bibitem[\protect\citeauthoryear{Grudi{\'c}, Guszejnov, Hopkins, Lamberts,
  {Boylan-Kolchin}, Murray  \& Schmitz}{Grudi{\'c}
  et~al.}{2018b}]{grudic_top_2018}
Grudi{\'c} M.~Y.,  Guszejnov D.,  Hopkins P.~F.,  Lamberts A.,
  {Boylan-Kolchin} M.,  Murray N.,   Schmitz D.,  2018b, \mn@doi [MNRAS]
  {10.1093/mnras/sty2303}, 481, 688

\bibitem[\protect\citeauthoryear{Grudi{\'c}, Hopkins, Lee, Murray,
  {Faucher-Gigu{\`e}re}  \& Johnson}{Grudi{\'c}
  et~al.}{2019}]{grudic_nature_2019}
Grudi{\'c} M.~Y.,  Hopkins P.~F.,  Lee E.~J.,  Murray N.,
  {Faucher-Gigu{\`e}re} C.-A.,   Johnson L.~C.,  2019, \mn@doi [MNRAS]
  {10.1093/mnras/stz1758}, 488, 1501

\bibitem[\protect\citeauthoryear{{H{\'e}nault-Brunet}
  et~al.,}{{H{\'e}nault-Brunet} et~al.}{2012}]{henault-brunet_vlt-flames_2012}
{H{\'e}nault-Brunet} V.,  et~al., 2012, \mn@doi [A\&A]
  {10.1051/0004-6361/201219472}, 545, L1

\bibitem[\protect\citeauthoryear{Heyer, Gutermuth, Urquhart, Csengeri, Wienen,
  Leurini, Menten  \& Wyrowski}{Heyer et~al.}{2016}]{heyer_rate_2016}
Heyer M.,  Gutermuth R.,  Urquhart J.~S.,  Csengeri T.,  Wienen M.,  Leurini
  S.,  Menten K.,   Wyrowski F.,  2016, \mn@doi [A\&A]
  {10.1051/0004-6361/201527681}, 588, A29

\bibitem[\protect\citeauthoryear{Howard, Pudritz  \& Harris}{Howard
  et~al.}{2018}]{howard_universal_2018}
Howard C.~S.,  Pudritz R.~E.,   Harris W.~E.,  2018, \mn@doi [Nat. Astron.]
  {10.1038/s41550-018-0506-0}, 2, 725

\bibitem[\protect\citeauthoryear{Kainulainen, Beuther, Henning  \&
  Plume}{Kainulainen et~al.}{2009}]{kainulainen_probing_2009}
Kainulainen J.,  Beuther H.,  Henning T.,   Plume R.,  2009, \mn@doi [A\&A]
  {10.1051/0004-6361/200913605}, 508, L35

\bibitem[\protect\citeauthoryear{Kamann et~al.,}{Kamann
  et~al.}{2018}]{kamann_stellar_2018}
Kamann S.,  et~al., 2018, \mn@doi [MNRAS] {10.1093/mnras/stx2719}, 473, 5591

\bibitem[\protect\citeauthoryear{Kamann, Bastian, Gieles, Balbinot  \&
  {H{\'e}nault-Brunet}}{Kamann et~al.}{2019}]{kamann_linking_2019}
Kamann S.,  Bastian N.~J.,  Gieles M.,  Balbinot E.,   {H{\'e}nault-Brunet} V.,
   2019, \mn@doi [MNRAS] {10.1093/mnras/sty3144}, 483, 2197

\bibitem[\protect\citeauthoryear{Kannan, Macci{\`o}, Fontanot, Moster, Karman
  \& Somerville}{Kannan et~al.}{2015}]{kannan_discs_2015}
Kannan R.,  Macci{\`o} A.~V.,  Fontanot F.,  Moster B.~P.,  Karman W.,
  Somerville R.~S.,  2015, \mn@doi [MNRAS] {10.1093/mnras/stv1633}, 452, 4347

\bibitem[\protect\citeauthoryear{Kim, Kim, Ostriker  \& Skinner}{Kim
  et~al.}{2017}]{kim_modeling_2017}
Kim J.-G.,  Kim W.-T.,  Ostriker E.~C.,   Skinner M.~A.,  2017, \mn@doi [ApJ]
  {10.3847/1538-4357/aa9b80}, 851, 93

\bibitem[\protect\citeauthoryear{Kim, Kim  \& Ostriker}{Kim
  et~al.}{2019}]{kim_modeling_2019}
Kim J.-G.,  Kim W.-T.,   Ostriker E.~C.,  2019, \mn@doi [ApJ]
  {10.3847/1538-4357/ab3d3d}, 883, 102

\bibitem[\protect\citeauthoryear{Kroupa}{Kroupa}{2001}]{kroupa_variation_2001}
Kroupa P.,  2001, \mn@doi [MNRAS] {10.1046/j.1365-8711.2001.04022.x}, 322, 231

\bibitem[\protect\citeauthoryear{Krumholz \& McKee}{Krumholz \&
  McKee}{2005}]{krumholz_general_2005}
Krumholz M.~R.,  McKee C.~F.,  2005, \mn@doi [ApJ] {10.1086/431734}, 630, 250

\bibitem[\protect\citeauthoryear{Krumholz \& Tan}{Krumholz \&
  Tan}{2007}]{krumholz_slow_2007}
Krumholz M.~R.,  Tan J.~C.,  2007, \mn@doi [ApJ] {10.1086/509101}, 654, 304

\bibitem[\protect\citeauthoryear{Krumholz, McKee  \& {Bland-Hawthorn}}{Krumholz
  et~al.}{2019}]{krumholz_star_2019}
Krumholz M.~R.,  McKee C.~F.,   {Bland-Hawthorn} J.,  2019, \mn@doi [ARA\&A]
  {10.1146/annurev-astro-091918-104430}, 57, 227

\bibitem[\protect\citeauthoryear{Lada \& Lada}{Lada \&
  Lada}{2003}]{lada_embedded_2003}
Lada C.~J.,  Lada E.~A.,  2003, \mn@doi [ARA\&A]
  {10.1146/annurev.astro.41.011802.094844}, 41, 57

\bibitem[\protect\citeauthoryear{Lah{\'e}n, Naab, Johansson, Elmegreen, Hu,
  Walch, Steinwandel  \& Moster}{Lah{\'e}n et~al.}{2020}]{lahen_griffin_2020}
Lah{\'e}n N.,  Naab T.,  Johansson P.~H.,  Elmegreen B.,  Hu C.-Y.,  Walch S.,
  Steinwandel U.~P.,   Moster B.~P.,  2020, \mn@doi [ApJ]
  {10.3847/1538-4357/ab7190}, 891, 2

\bibitem[\protect\citeauthoryear{Lardo et~al.,}{Lardo
  et~al.}{2015}]{lardo_gaia_2015}
Lardo C.,  et~al., 2015, \mn@doi [A\&A] {10.1051/0004-6361/201425036}, 573,
  A115

\bibitem[\protect\citeauthoryear{Larson}{Larson}{1969}]{larson_numerical_1969}
Larson R.~B.,  1969, \mn@doi [MNRAS] {10.1093/mnras/145.3.271}, 145, 271

\bibitem[\protect\citeauthoryear{Lee, Chang  \& Murray}{Lee
  et~al.}{2015}]{lee_time-varying_2015}
Lee E.~J.,  Chang P.,   Murray N.,  2015, \mn@doi [ApJ]
  {10.1088/0004-637X/800/1/49}, 800, 49

\bibitem[\protect\citeauthoryear{Lee, {Miville-Desch{\^e}nes}  \& Murray}{Lee
  et~al.}{2016}]{lee_observational_2016}
Lee E.~J.,  {Miville-Desch{\^e}nes} M.-A.,   Murray N.~W.,  2016, \mn@doi [ApJ]
  {10.3847/1538-4357/833/2/229}, 833, 229

\bibitem[\protect\citeauthoryear{Li}{Li}{2018}]{li_scale-free_2018}
Li G.-X.,  2018, \mn@doi [MNRAS] {10.1093/mnras/sty657}, 477, 4951

\bibitem[\protect\citeauthoryear{Li \& Gnedin}{Li \&
  Gnedin}{2019}]{li_star_2019}
Li H.,  Gnedin O.~Y.,  2019, \mn@doi [MNRAS] {10.1093/mnras/stz1114}, 486, 4030

\bibitem[\protect\citeauthoryear{Li, Gnedin, Gnedin, Meng, Semenov  \&
  Kravtsov}{Li et~al.}{2017}]{li_star_2017}
Li H.,  Gnedin O.~Y.,  Gnedin N.~Y.,  Meng X.,  Semenov V.~A.,   Kravtsov
  A.~V.,  2017, \mn@doi [ApJ] {10.3847/1538-4357/834/1/69}, 834, 69

\bibitem[\protect\citeauthoryear{Li, Vogelsberger, Marinacci  \& Gnedin}{Li
  et~al.}{2019}]{li_disruption_2019}
Li H.,  Vogelsberger M.,  Marinacci F.,   Gnedin O.~Y.,  2019, \mn@doi [MNRAS]
  {10.1093/mnras/stz1271}, 487, 364

\bibitem[\protect\citeauthoryear{Lombardi, Lada  \& Alves}{Lombardi
  et~al.}{2008}]{lombardi_2mass_2008}
Lombardi M.,  Lada C.~J.,   Alves J.,  2008, \mn@doi [A\&A]
  {10.1051/0004-6361:200810070}, 489, 143

\bibitem[\protect\citeauthoryear{Lucas, Bonnell  \& Dale}{Lucas
  et~al.}{2020}]{lucas_supernova_2020}
Lucas W.~E.,  Bonnell I.~A.,   Dale J.~E.,  2020, \mn@doi [MNRAS]
  {10.1093/mnras/staa451}, 493, 4700

\bibitem[\protect\citeauthoryear{Ma et~al.,}{Ma
  et~al.}{2020}]{ma_self-consistent_2020}
Ma X.,  et~al., 2020, \mn@doi [MNRAS] {10.1093/mnras/staa527}, 493, 4315

\bibitem[\protect\citeauthoryear{Mackey, Da~Costa, Ferguson  \& Yong}{Mackey
  et~al.}{2013}]{mackey_vlt/flames_2013}
Mackey A.~D.,  Da~Costa G.~S.,  Ferguson A. M.~N.,   Yong D.,  2013, \mn@doi
  [ApJ] {10.1088/0004-637X/762/1/65}, 762, 65

\bibitem[\protect\citeauthoryear{Mapelli}{Mapelli}{2017}]{mapelli_rotation_2017}
Mapelli M.,  2017, \mn@doi [MNRAS] {10.1093/mnras/stx304}, 467, 3255

\bibitem[\protect\citeauthoryear{Marinacci, Pakmor  \& Springel}{Marinacci
  et~al.}{2014}]{marinacci_formation_2014}
Marinacci F.,  Pakmor R.,   Springel V.,  2014, \mn@doi [MNRAS]
  {10.1093/mnras/stt2003}, 437, 1750

\bibitem[\protect\citeauthoryear{Martig, Bournaud, Croton, Dekel  \&
  Teyssier}{Martig et~al.}{2012}]{martig_diversity_2012}
Martig M.,  Bournaud F.,  Croton D.~J.,  Dekel A.,   Teyssier R.,  2012,
  \mn@doi [ApJ] {10.1088/0004-637X/756/1/26}, 756, 26

\bibitem[\protect\citeauthoryear{McKee \& Ostriker}{McKee \&
  Ostriker}{2007}]{mckee_theory_2007}
McKee C.~F.,  Ostriker E.~C.,  2007, \mn@doi [ARA\&A]
  {10.1146/annurev.astro.45.051806.110602}, 45, 565

\bibitem[\protect\citeauthoryear{Mueller, Shirley, Evans~II  \&
  Jacobson}{Mueller et~al.}{2002}]{mueller_physical_2002}
Mueller K.~E.,  Shirley Y.~L.,  Evans~II N.~J.,   Jacobson H.~R.,  2002,
  \mn@doi [ApJS] {10.1086/342881}, 143, 469

\bibitem[\protect\citeauthoryear{Murray \& Chang}{Murray \&
  Chang}{2015}]{murray_star_2015}
Murray N.,  Chang P.,  2015, \mn@doi [ApJ] {10.1088/0004-637X/804/1/44}, 804,
  44

\bibitem[\protect\citeauthoryear{Murray, Quataert  \& Thompson}{Murray
  et~al.}{2010}]{murray_disruption_2010}
Murray N.,  Quataert E.,   Thompson T.~A.,  2010, \mn@doi [ApJ]
  {10.1088/0004-637X/709/1/191}, 709, 191

\bibitem[\protect\citeauthoryear{Murray, Chang, Murray  \& Pittman}{Murray
  et~al.}{2017}]{murray_collapse_2017}
Murray D.~W.,  Chang P.,  Murray N.~W.,   Pittman J.,  2017, \mn@doi [MNRAS]
  {10.1093/mnras/stw2796}, 465, 1316

\bibitem[\protect\citeauthoryear{Myers, Klein, Krumholz  \& McKee}{Myers
  et~al.}{2014}]{myers_star_2014}
Myers A.~T.,  Klein R.~I.,  Krumholz M.~R.,   McKee C.~F.,  2014, \mn@doi
  [MNRAS] {10.1093/mnras/stu190}, 439, 3420

\bibitem[\protect\citeauthoryear{{Naranjo-Romero}, {V{\'a}zquez-Semadeni}  \&
  Loughnane}{{Naranjo-Romero} et~al.}{2015}]{naranjo-romero_hierarchical_2015}
{Naranjo-Romero} R.,  {V{\'a}zquez-Semadeni} E.,   Loughnane R.~M.,  2015,
  \mn@doi [ApJ] {10.1088/0004-637X/814/1/48}, 814, 48

\bibitem[\protect\citeauthoryear{Obreja, Stinson, Dutton, Macci{\`o}, Wang  \&
  Kang}{Obreja et~al.}{2016}]{obreja_nihao_2016}
Obreja A.,  Stinson G.~S.,  Dutton A.~A.,  Macci{\`o} A.~V.,  Wang L.,   Kang
  X.,  2016, \mn@doi [MNRAS] {10.1093/mnras/stw690}, 459, 467

\bibitem[\protect\citeauthoryear{Ostriker, Stone  \& Gammie}{Ostriker
  et~al.}{2001}]{ostriker_density_2001}
Ostriker E.~C.,  Stone J.~M.,   Gammie C.~F.,  2001, \mn@doi [ApJ]
  {10.1086/318290}, 546, 980

\bibitem[\protect\citeauthoryear{Padoan, Haugb{\o}lle  \& Nordlund}{Padoan
  et~al.}{2012}]{padoan_simple_2012}
Padoan P.,  Haugb{\o}lle T.,   Nordlund {\AA}.,  2012, \mn@doi [ApJ]
  {10.1088/2041-8205/759/2/L27}, 759, L27

\bibitem[\protect\citeauthoryear{Palau et~al.,}{Palau
  et~al.}{2014}]{palau_fragmentation_2014}
Palau A.,  et~al., 2014, \mn@doi [ApJ] {10.1088/0004-637X/785/1/42}, 785, 42

\bibitem[\protect\citeauthoryear{Penston}{Penston}{1969}]{penston_dynamics_1969}
Penston M.~V.,  1969, \mn@doi [MNRAS] {10.1093/mnras/144.4.425}, 144, 425

\bibitem[\protect\citeauthoryear{Pfeffer, Kruijssen, Crain  \& Bastian}{Pfeffer
  et~al.}{2018}]{pfeffer_e-mosaics_2018}
Pfeffer J.,  Kruijssen J. M.~D.,  Crain R.~A.,   Bastian N.,  2018, \mn@doi
  [MNRAS] {10.1093/mnras/stx3124}, 475, 4309

\bibitem[\protect\citeauthoryear{Pirogov}{Pirogov}{2009}]{pirogov_density_2009}
Pirogov L.~E.,  2009, \mn@doi [Astron. Rep.] {10.1134/S1063772909120051}, 53,
  1127

\bibitem[\protect\citeauthoryear{Raskutti, Ostriker  \& Skinner}{Raskutti
  et~al.}{2016}]{raskutti_numerical_2016}
Raskutti S.,  Ostriker E.~C.,   Skinner M.~A.,  2016, \mn@doi [ApJ]
  {10.3847/0004-637X/829/2/130}, 829, 130

\bibitem[\protect\citeauthoryear{Rogers \& Pittard}{Rogers \&
  Pittard}{2013}]{rogers_feedback_2013}
Rogers H.,  Pittard J.~M.,  2013, \mn@doi [MNRAS] {10.1093/mnras/stt255}, 431,
  1337

\bibitem[\protect\citeauthoryear{Schneider et~al.,}{Schneider
  et~al.}{2013}]{schneider_what_2013}
Schneider N.,  et~al., 2013, \mn@doi [ApJ] {10.1088/2041-8205/766/2/L17}, 766,
  L17

\bibitem[\protect\citeauthoryear{Schneider et~al.,}{Schneider
  et~al.}{2015}]{schneider_understanding_2015}
Schneider N.,  et~al., 2015, \mn@doi [A\&A] {10.1051/0004-6361/201423569}, 575,
  A79

\bibitem[\protect\citeauthoryear{Scoville \& Good}{Scoville \&
  Good}{1989}]{scoville_far-infrared_1989}
Scoville N.~Z.,  Good J.~C.,  1989, \mn@doi [ApJ] {10.1086/167283}, 339, 149

\bibitem[\protect\citeauthoryear{Shu, Adams  \& Lizano}{Shu
  et~al.}{1987}]{shu_star_1987}
Shu F.~H.,  Adams F.~C.,   Lizano S.,  1987, \mn@doi [ARA\&A]
  {10.1146/annurev.aa.25.090187.000323}, 25, 23

\bibitem[\protect\citeauthoryear{Skinner \& Ostriker}{Skinner \&
  Ostriker}{2015}]{skinner_numerical_2015}
Skinner M.~A.,  Ostriker E.~C.,  2015, \mn@doi [ApJ]
  {10.1088/0004-637X/809/2/187}, 809, 187

\bibitem[\protect\citeauthoryear{Sokolowska, Capelo, Fall, Mayer, Shen  \&
  Bonoli}{Sokolowska et~al.}{2017}]{sokolowska_galactic_2017}
Sokolowska A.,  Capelo P.~R.,  Fall S.~M.,  Mayer L.,  Shen S.,   Bonoli S.,
  2017, \mn@doi [ApJ] {10.3847/1538-4357/835/2/289}, 835, 289

\bibitem[\protect\citeauthoryear{Springel}{Springel}{2010}]{springel_e_2010}
Springel V.,  2010, \mn@doi [MNRAS] {10.1111/j.1365-2966.2009.15715.x}, 401,
  791

\bibitem[\protect\citeauthoryear{Springel, White, Tormen  \&
  Kauffmann}{Springel et~al.}{2001}]{springel_populating_2001}
Springel V.,  White S. D.~M.,  Tormen G.,   Kauffmann G.,  2001, \mn@doi
  [MNRAS] {10.1046/j.1365-8711.2001.04912.x}, 328, 726

\bibitem[\protect\citeauthoryear{Tiongco, Vesperini  \& Varri}{Tiongco
  et~al.}{2018}]{tiongco_complex_2018}
Tiongco M.~A.,  Vesperini E.,   Varri A.~L.,  2018, \mn@doi [MNRAS]
  {10.1093/mnrasl/sly009}, 475, L86

\bibitem[\protect\citeauthoryear{Vutisalchavakul, Evans~II  \&
  Heyer}{Vutisalchavakul et~al.}{2016}]{vutisalchavakul_star_2016}
Vutisalchavakul N.,  Evans~II N.~J.,   Heyer M.,  2016, \mn@doi [ApJ]
  {10.3847/0004-637X/831/1/73}, 831, 73

\bibitem[\protect\citeauthoryear{Walch, Whitworth, Bisbas, W{\"u}nsch  \&
  Hubber}{Walch et~al.}{2012}]{walch_dispersal_2012}
Walch S.~K.,  Whitworth A.~P.,  Bisbas T.,  W{\"u}nsch R.,   Hubber D.,  2012,
  \mn@doi [MNRAS] {10.1111/j.1365-2966.2012.21767.x}, 427, 625

\bibitem[\protect\citeauthoryear{Wu, Evans, Shirley  \& Knez}{Wu
  et~al.}{2010}]{wu_properties_2010}
Wu J.,  Evans N.~J.,  Shirley Y.~L.,   Knez C.,  2010, \mn@doi [ApJS]
  {10.1088/0067-0049/188/2/313}, 188, 313

\bibitem[\protect\citeauthoryear{Wyrowski et~al.,}{Wyrowski
  et~al.}{2016}]{wyrowski_infall_2016}
Wyrowski F.,  et~al., 2016, \mn@doi [A\&A] {10.1051/0004-6361/201526361}, 585,
  A149

\bibitem[\protect\citeauthoryear{Zasov \& Zaitseva}{Zasov \&
  Zaitseva}{2017}]{zasov_hi_2017}
Zasov A.~V.,  Zaitseva N.~A.,  2017, \mn@doi [Astron. Lett.]
  {10.1134/S1063773717070052}, 43, 439

\bibitem[\protect\citeauthoryear{Zuckerman \& Evans}{Zuckerman \&
  Evans}{1974}]{zuckerman_models_1974}
Zuckerman B.,  Evans Ii N.~J.,  1974, \mn@doi [ApJ] {10.1086/181613}, 192, L149

\makeatother
\end{thebibliography}




\appendix

\section{Test of different star formation density thresholds}
\label{append:threshold}

\begin{figure}
\includegraphics[width=\columnwidth]{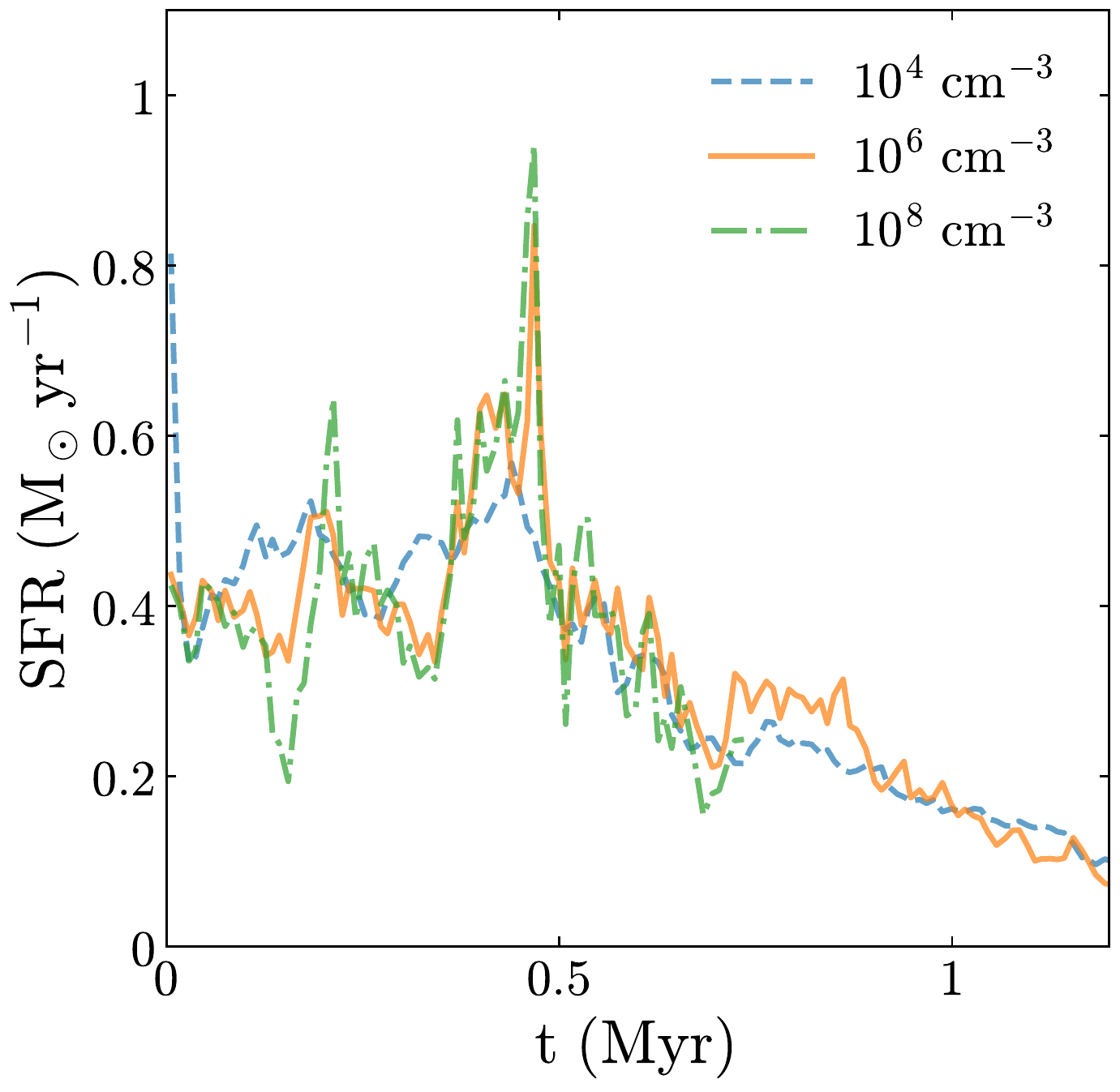}
\caption{SFR evolution of a PL20 run with different star formation density thresholds varying from $n_\mathrm{cell}=10^4$ to $10^8\ \mathrm{cm^{-3}}$, where the SFR is calculated as the time derivative of total stellar mass. }
  \label{fig:thresholdAppend}
\end{figure}

As mentioned in Sec.~\ref{sec:methods}, the star forming cells are identified as cold, contracting, self-gravitating and dense gas cells. We define the dense gas cells as cells with density above the star formation density threshold, $n_\mathrm{cell}$. Using a PL20 run as an example, we test three thresholds of $n_\mathrm{cell}=10^4$, $10^6$, and $10^8\ \mathrm{cm^{-3}}$, and plot the SFR evolution in Fig.~\ref{fig:thresholdAppend}. 
The PL20 run with $n_\mathrm{cell}=10^4\ \mathrm{cm^{-3}}$ triggers a high SFR at the very beginning because a large amount of gas near the center is denser than this threshold (see Fig.~\ref{fig:iniPro}). These cells are immediately converted into stars at the first few timesteps of the simulation, leading to an artificial starburst that depends strongly on the choice of the threshold. We thus recommend $n_\mathrm{cell}>10^5\ \mathrm{cm^{-3}}$ to fully resolve the initial star formation. We also find that the thresholds of $n_\mathrm{cell}=10^6$ and $10^8\ \mathrm{cm^{-3}}$ do not affect the SFH significantly. However, the $n_\mathrm{cell}=10^8\ \mathrm{cm^{-3}}$ simulations is computationally expensive because it requires finer resolution. Thus, we use $n_\mathrm{cell}=10^6~\mathrm{cm^{-3}}$ in this work to both avoid a numerical bias and save computing time.

\section{Calculation of accretion rate}
\label{append:accretionRate}

In Sec.~\ref{sec:evolutionOfGasDensityProfile}, we propose a method to determine the accretion rate profiles of GMCs. Here, we compare this method (or ``method 1'' hereafter) with another method in \citet[][or ``method 2'' hereafter]{howard_universal_2018} as below.

\begin{itemize}
    \item Method 1: we firstly count all Voronoi cells intersecting the sphere with radius of $r_0$. Then, we define the accretion rate of the $n$-th intersecting cell as $\dot{m}_n=u_n A_n\rho_n$, where $A_n$ is the intersecting area, $\rho_n$ is the density, and $u_n$ is the incoming velocity, which is the compressive component of the velocity $\mathbf{v}_n$. Since calculating the exact value of $A_n$ is computationally expensive, we apply a simplified method to approximate $A_n$: we first consider the $n$-th intersecting cell as a sphere whose effective radius is $r_n=(3V_n/4\pi)^{1/3}$, where $V_n=m_n/\rho_n$ is the volume of this cell; we then compute $A_n$ as the area of the $r_0$ sphere intersecting this cell, which is
    \begin{equation}
    	A_n=2\pi r_0^2\left(1-\frac{l_n^2+r_0^2-r_n^2}{2r_0l_n}\right),
    	\label{eq:intersect}
    \end{equation}
    where $l_n$ is the spatial separation between this cell and the center. Now that we have a good definition for $A_n$, we can calculate the accretion rate as the summation of $\dot{m}_n$, i.e.,
    \begin{equation}
        \dot{M}^i(r_0)=\Sigma_n\dot{m}_n=\Sigma_n u^i_n A^i_n \rho^i_n,
        \label{eq:accretionMethod2}
    \end{equation}
    in which the label $i$ refers to the $i$-th snapshot. It is worth notifying that the summation of $A_n$ is expected to be $4\pi r_0^2$, but Eq.~\ref{eq:intersect} does not guarantee such expectation. Therefore, we multiply Eq.~\ref{eq:accretionMethod2} by $4\pi r_0^2/\Sigma_n A_n \rho^i_n$ as the final accretion rate.
    \item Method 2: the accretion rate at radius $r_0$ is calculated as
    \begin{equation}
        \dot{M}^i(r_0)=\frac{M^i_r-M^{i-1}_r}{t_i-t_{i-1}},
        \label{eq:accretionMethod1}
    \end{equation}
    where $M^i(r_0)$ represents the total mass enclosed in the radius $r_0$. This method requires two adjacent snapshots and computes $\dot{M}^i(r_0)$ as the average accretion rate between them.
\end{itemize}

\begin{figure}
\includegraphics[width=\columnwidth]{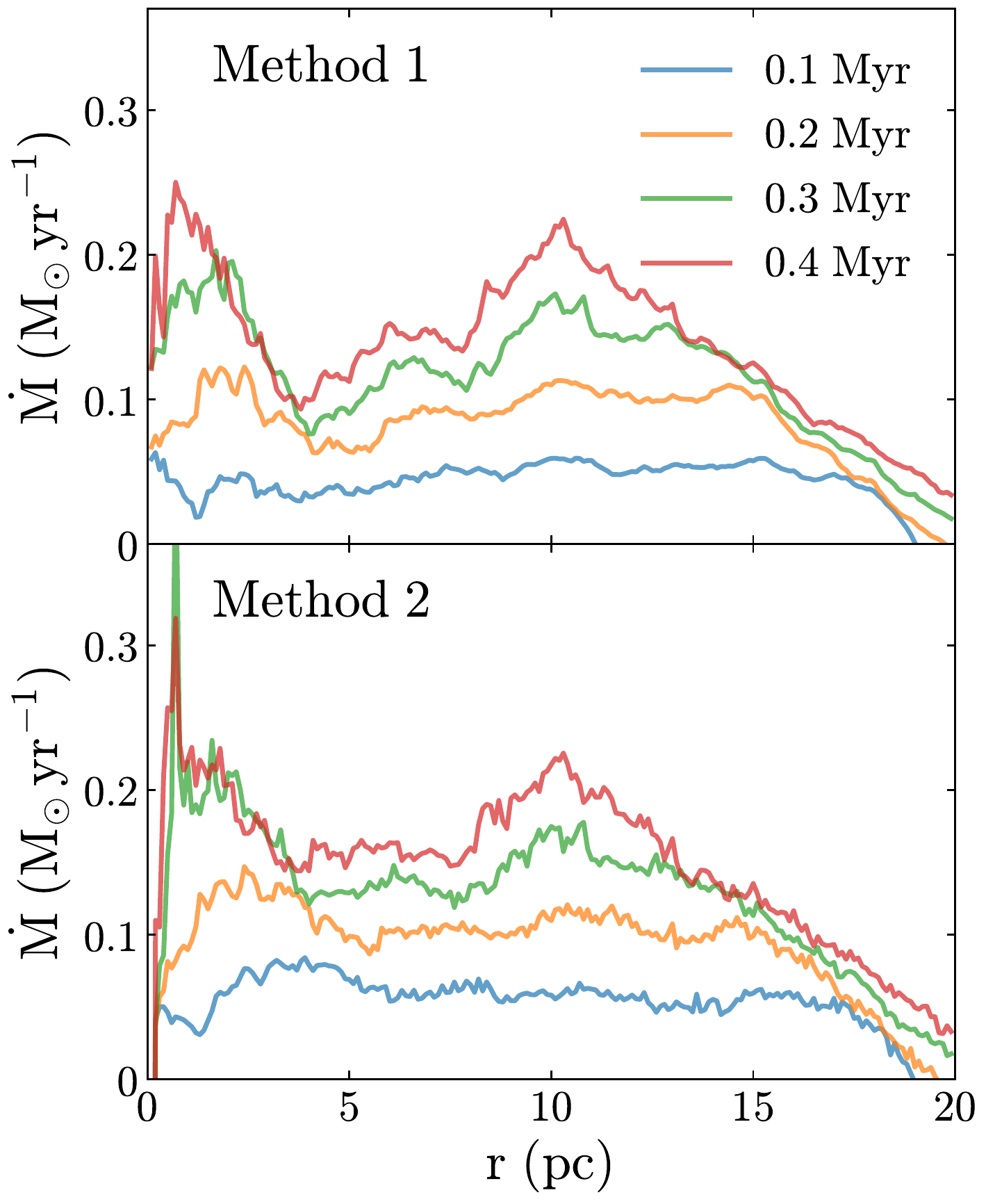}
\caption{Comparison of accretion rate profiles of a PL15 run determined by our method (top) and the method in \citet[][bottom]{howard_universal_2018} at different times from $t=0.1$ to $0.4\ \mathrm{Myr}$.}
  \label{fig:accretionAppend}
\end{figure}

Methods 1 and 2 correspond to the first and second terms of the integral form of continuity equation:
\begin{equation}
    \oiint_{r=r_0}\rho\mathbf{v}\cdot\mathbf{n}\,d\sigma + \frac{\partial}{\partial t}\iiint_{r<r_0}\rho\,d^3\mathbf{r}=0.
    \label{eq:continuity}
\end{equation}
Thus, the two methods are equivalent in an analytic view. Next, we compare the two methods numerically. Using a PL15 run as an example, we plot the accretion rate profiles at different time in Fig.~\ref{fig:accretionAppend} to compare the numerical performance of the two methods. Although the two methods show no systematic difference, we consider method 1 a better choice here for three reasons: first, method 1 produces smoother curves than method 2; second, method 1 can give the instantaneous accretion rate at the epoch of the snapshot, while method 2 only produces the averaged accretion rate between two snapshots; third, method 1 can be easily implemented in the moving mesh code \textsc{Arepo}. Therefore, we use method 1 to determine the accretion rate in this work.

Because of the initial ($t=0$) turbulent field, methods 1 and 2 both produce nonzero values when implemented for the initial profiles. These initial values are completely random, leading to large variation of accretion rate profiles. To minimize the influence of such randomness, we add an additional step to the above methods: subtracting the accretion rate profile at $t=0$ from those at $t\neq 0$.

\bsp	
\label{lastpage}
\end{document}